\documentclass[12pt,a4paper]{iopart}

\usepackage{iopams}
\usepackage{graphicx}
\usepackage{hyperref}

\begin{document}

\review[Micrometer-scale refrigerators]{Micrometer-scale refrigerators}
\author{Juha T. Muhonen}
\address{Low Temperature Laboratory, Aalto University, School of Science, P.O. Box 13500, 00076 Aalto, Finland}
\address{Department of Physics, University of Warwick, Coventry CV4 7AL, United Kingdom}
\author{Matthias Meschke}
\address{Low Temperature Laboratory, Aalto University, School of Science, P.O. Box 13500, 00076 Aalto, Finland}
\author{Jukka P. Pekola}
\address{Low Temperature Laboratory, Aalto University, School of Science, P.O. Box 13500, 00076 Aalto, Finland}
\ead{jukka.pekola@aalto.fi}

\begin{abstract}
A superconductor with a gap in the density of states or a quantum dot with discrete energy levels is a central building block in realizing an electronic on-chip cooler. They can work as energy filters, allowing only hot quasi-particles to tunnel out from the electrode to be cooled. This principle has been employed experimentally since the early 1990's in investigations and demonstrations of micron-scale coolers at sub-kelvin temperatures. In this paper, we review the basic experimental conditions in realizing the coolers and the main practical issues that are known to limit their performance. We give an update of experiments performed on cryogenic micron-scale coolers in the past five years.
\end{abstract}


\maketitle

\section{Introduction}

Electron transport in micro- and nano-structures has attracted lots of attention over the past several decades. Until recently, less concern has been paid on the associated energy currents and generation of heat. However, heat currents and dissipation often limit the performance of an electronic device in particular at cryogenic temperatures. Cooling a device to lower temperatures generally improves its characteristics in terms of increased sensitivity and decreased noise. Despite the fast progress in liquid cryogen free cooling techniques, refrigeration to cryogenic temperatures remains expensive and proper infrastructure for cryogenic work is found typically only in specialized laboratories. Therefore, it is of interest to explore cooling techniques that operate directly on a chip, even though they may be an option only in special applications (such as direct detector cooling with limited cooling power). They could provide an alternative solution as the final stage of a refrigerator which is both economic and easy-to-use.

\begin{figure}
\centering
\includegraphics{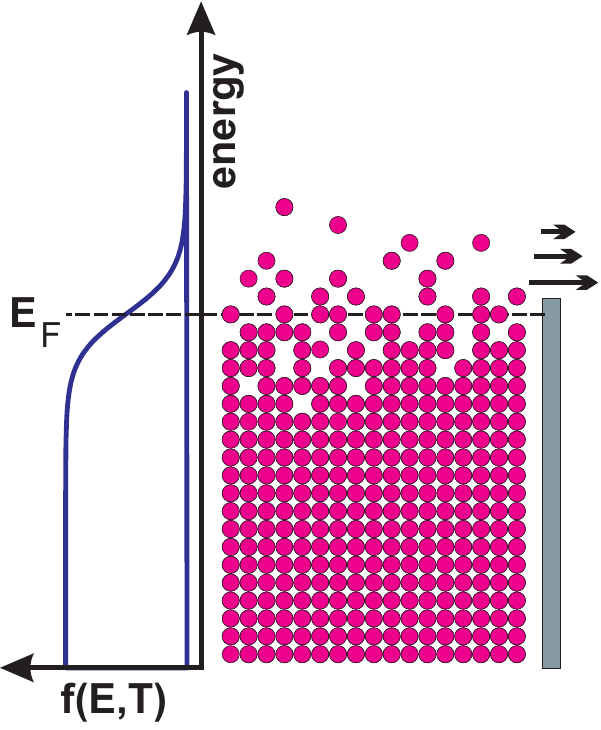}
\caption{The basic principle of direct eletronic cooling. An energy filter (gray wall) allows only high energy electrons (red circles) to be removed from the electron system. This ejection leads to sharpening of the electron distribution, i.e. cooling. The corresponding Fermi distribution is shown on the left. The density of states of the conductor is to a good approximation constant on the narrow energy range of interest, as the thermal energy $k_BT$ is small compared to the Fermi energy.
}
\label{fig:zero}
\end{figure}

In this review, we will concentrate on low temperature electronic on-chip coolers. The basic principle of operation is shown in \fref{fig:zero}. An energy filter allows only high energy electrons to be removed from an electron system, hence leading to cooling of the system. One possible energy filter is the superconducting gap. Experimental activity using on-chip cooling by NIS (N = normal conductor, I = insulator, S = superconductor) junctions has a history of less than twenty years now. We focus on progress over the past five years as the earlier achievements have been extensively covered in another review \cite{Giazotto2006}. To our knowledge, only a handful of laboratories are actively experimenting on electronic coolers at sub-kelvin temperatures at the current time. This has made our task of covering most of the published activity on the topic hopefully successful. The aim in the present review is to report mainly on experiments and the associated phenomena with less emphasis on the theoretical background. We do discuss various energy relaxation mechanisms in a cooler to some length since their role has turned out to be centrally important in trying to achieve optimum performance of an electronic cooler. For instance, the quasi-particle relaxation in a superconductor at the back-side of the cooler is a major, still unresolved problem limiting the performance of a NIS refrigerator. This is a particularly important issue when low temperatures and enhanced cooling powers are to be achieved. After discussing heat transport and dissipation, we review recent conceptual and technological advances in terms of cooling principles, materials and physical realizations. Although most of the reported work deals with superconductor-based solutions, we want to note already here that a relatively recent experiment using quantum dots as energy filters in semiconducting 2DEG structures \cite{Prance2009} has spurred research in new potential realizations of practical on-chip coolers.
\section{Temperatures and energy relaxation}

\subsection{Temperatures of a micron-scale conductor}

It is not trivial to define the temperature of a micron-scale conductor at sub-kelvin temperatures. First, one does not have just one system but an ensemble of subsystems, which each have a characteristic internal energy and are coupled to each other very non-linearly. The most relevant subsystems concerning the micron-scale refrigerators discussed here are the electron system, the phonon system of the conductor, and the phonon system of the substrate. Secondly, in order for temperature to be a well defined concept, the relaxation rates inside each subsystem must be faster than the couplings between them. Then the systems will follow thermal distributions (Fermi-Dirac for electrons, Bose-Einstein for phonons) and temperatures can be related to them. The situation where this is true but the effective temperatures of different subsystems are not equal, is generally called \textit{quasi-equilibrium}. The situation where energy is exchanged with the system faster than it can relax and hence no temperature can be defined for it is called \textit{non-equilibrium}. Full \textit{equilibrium} would be the situation where all the subsystems are at the same temperature.

\begin{figure}
\centering
\includegraphics{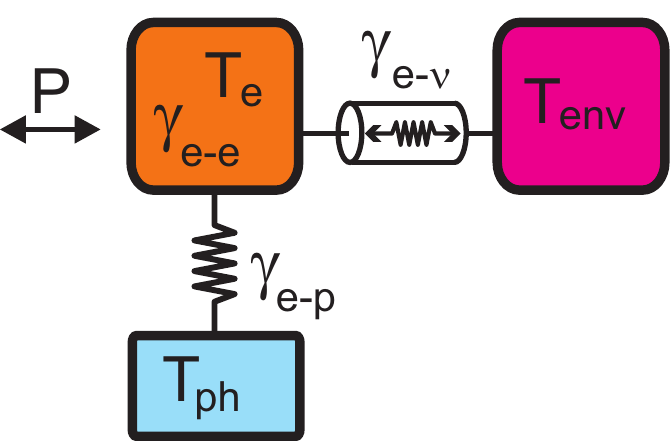}
\caption{Simplified thermal model of a conductor at temperature $T_e$ (S or N). External power $P$ is exchanged with the system. Electron-electron ($\gamma _{e-e}$) interaction drives the electron subsystem towards a quasi-equilibrium distribution and electron-phonon interaction couples it to the phonon bath ($\gamma _{e-p}$). It is also coupled electromagnetically to the environment (at temperature $T_{env}$) which can be spatially well separated from the cooled volume.
}
\label{fig:thmodel}
\end{figure}

In what follows, we will be mainly concerned with the electron system, as the micron-scale refrigerators discussed in this review cool the electron system directly. \Fref{fig:thmodel} gives a simplified thermal model of a conductor, S or N. The basic picture of cooling, described in the following chapters, only holds if the electron system stays in quasi-equilibrium. For this to be true, the electron-electron (e-e) collision rate $\gamma_{e-e}$ must be faster than the injection rate of quasiparticles and photons. Then the distribution can be described by a Fermi-Dirac one and we can ascribe a well defined temperature to the subsystem. This is the prevailing situation, and it has turned out to be difficult to overcome this in tunnel-coupled systems, particularly in the N electrodes, see however \cite{Pekola2004}. 
In a system where electrons can be injected directly without a tunnel barrier, non-equilibrium distributions have been observed in several experiments (e.g. the seminal paper by Pothier \textit{et al.} \cite{Pothier1997}). On the other hand, the ratio between the $e-p$ collision rate $\gamma_{e-p}$ and the injection rate (or the photon exchange rate $\gamma_{e-\nu}$) determines whether the electron sub-system has the same temperature as the phonon bath or not. Low $\gamma_{e-p}$ means that the electronic system can have a temperature different from that of the bath which makes the direct electronic cooling possible. This happens in particular at low temperatures since the relaxation gets increasingly slow at low temperatures (e.g. for normal metals $\gamma_{e-p}\propto T^{-3}$). If the coupling to the phonon system is suppressed, the photonic coupling to the environment $\gamma_{e-\nu}$ can become the dominant relaxation mechanism as will be discussed later in the review.

The phonon systems in the conductor and in the substrate are also in principle two separate systems which can have differing temperatures. As a result of lattice mismatch between the two materials, the phonons can be scattered at the interface leading to thermal resistance, known as Kapitza resistance. However, at the low temperatures considered here, the dominant wavelength of the thermal phonons is of the order of several micrometers which is much larger than the thickness of a typical metallic or semiconducting film. Hence the interface (and difference between the two materials) should be quite transparent to these phonons. In addition, as the electron-phonon coupling (discussed below) decreases rapidly at low temperatures, this thermal resistance will be the dominant thermal bottleneck. (See, however, \cite{Rajauria2007}.) We will neglect the Kapitza resistance throughout.

\subsection{Relaxation mechanisms at low temperatures}

\subsubsection{Electrons in metals and semiconductors}

In most cases encountered experimentally, the electron-phonon coupling is the dominant inelastic scattering mechanism for the electron system at not too low temperatures. The electron-phonon relaxation in ordinary metals has been discussed and measured in various experiments over the past several decades. With the quasi-equilibrium conditions, a straightforward first order perturbation theory calculation (assuming scalar deformation-potential coupling and three-dimensional electron and phonon systems) yields an energy exchange rate \cite{Wellstood1994}
\begin{equation} \label{eq:eph}
P_{e-p}^n (T_e, T_p) = \Sigma \mathcal{V}(T_e^5-T_p^5).
\end{equation}
Here $\Sigma$ is the material parameter, known for most ordinary metals, $\mathcal{V}$ is the volume of the conductor, $T_e$ and $T_p$ are the temperatures of the electron and phonon system respectively. Substantial deviations from this law, which is obeyed experimentally astonishingly well irrespective of the particular normal metal material or geometry, are expected in restricted dimensions, for superconductors, and for semiconductors. We discuss these issues below. For most metals, one has $\Sigma \sim 10^9$ WK$^{-5}$m$^{-3}$.

In semiconductors, the situation is different from the normal metal case essentially because of two facts: the small amount of momentum-space that is occupied in the semiconductor and the presence of the band gap. In essence, the coupling between phonons and the electron system can then be described with a deformation potential constant that describes how the minimum of the conduction band moves in response to the stresses caused by phonons. Hence, variables of the electron system (mainly the momentum distribution) can be neglected. We delay more detailed discussion about this issue to the section about Schottky coolers.

So far, we have implicitly assumed that the power is distributed uniformly on the conductor or that the conductor has a high enough thermal conductivity so that no temperature gradients exist inside it. In practice this is often not the case. Especially considering the micron-scaled coolers, the prevailing situation is such that one has a pointlike cooling/heating source on one end of a conducting wire or a plate. To make an accurate model in such situations, it becomes compulsory to consider also the thermal conductivity inside the electron system. If one makes the assumptions outlined above (so that a position dependent temperature can be defined), then the thermal conductivity in normal metals at low temperatures follows textbook models of the electron gas. The heat current density is related to the temperature gradient as
$Q = -\kappa_n \nabla T$, 
where $\kappa_n$ is the thermal conductivity. This can be related to the electrical conductivity $\sigma$ via Wiedemann-Franz law $\kappa_n = \mathcal{L}\sigma T$, where $\mathcal{L}$ is the Lorentz number. With these assumptions, a steady-state diffusion equation can be written for a differential volume element
\begin{equation}
\nabla \cdot [-\kappa_n(T_e,x) \nabla T_e(x)] = \Sigma [T_e(x)^5-T_p(x)^5] + \mathcal{P}_{\rm ext}(x),
\label{eq:diffN}
\end{equation}
where we have used $\mathcal{P}_{\rm ext}$ as the power density from all possible external heating sources. Solving this equation self-consistently and with proper boundary conditions will yield the temperature profile of the conductor.

\subsubsection{Quasiparticle excitations in superconductors} \label{sec:qps}

In a superconductor, we are interested in the system of quasiparticle excitations. The Cooper pair condensate carries no entropy and has no explicit role in the thermal properties discussed here. The typically dominant relaxation mechanisms are analogous to the normal metal case: the quasiparticle heat conductivity along the wire and quasiparticle-phonon relaxation (which is determined predominantly by the recombination of quasiparticles into Cooper pairs). The most obvious differences to the normal metal case are: (i) the exponentially small amount of quasiparticles at temperatures $T \ll T_C$ and (ii) the fact that the quasiparticles need to absorb or emit energy larger than the superconducting gap $\Delta$. Combining these effects leads to exponentially suppressed heat conductivity and relaxation at low temperatures [$\kappa,\Sigma \propto \exp(-\Delta/(k_BT))$].

\begin{figure}
\centering
\includegraphics{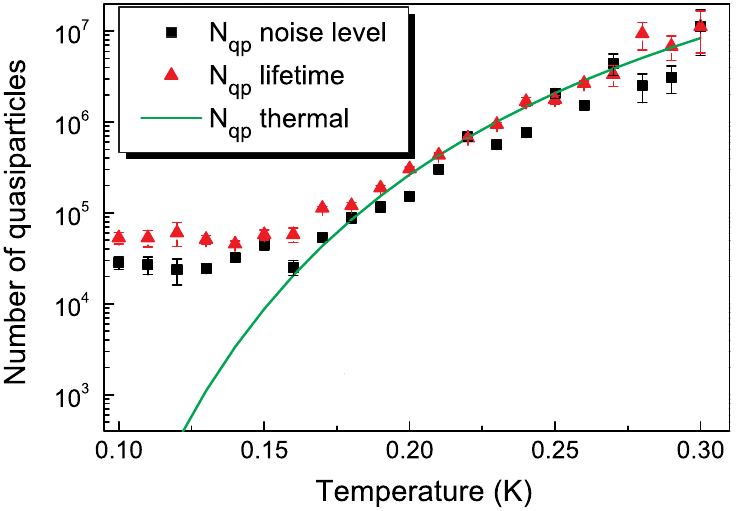}
\caption{Theoretical (line) and experimental (dots) quasiparticle density at low temperatures. The low temperature anomaly can be clearly seen. Although the saturation level depends on the particular experimental set-up, the same qualitative behaviour has been seen in all experiments. 
\textit{Reprinted figure with permission from \cite{Visser2011}. Copyright 2011 by the American Physical Society.}
}
\label{fig:qps}
\end{figure}

Ideally, a superconductor at a temperature well below $T_C$ has a negligible number of quasiparticle excitations. Quantitatively, the BCS-theory predicts the quasiparticle density $n_{\rm qp}$ in thermal equilibrium to be
\begin{equation} \label{eq:thqp}
n_{\rm qp} = 2 N_F \int_\Delta^\infty \rmd E \frac{E}{\sqrt{E^2-\Delta^2}} f(E)	\approx N_F \sqrt{2 \pi k_BT\Delta}\,e^{-\Delta/k_BT},
\end{equation}
where $N_F$ is the density of states at the Fermi level (in the normal state), $f$ is the Fermi distribution function. The last step applies for $\Delta/(k_BT) \gg 1$. The factor of 2 comes from the fact that we should also integrate over the negative energies (holelike quasiparticles). For illustration, one can put the parameters of aluminium (Al) to \eref{eq:thqp} ($\Delta/k_B =2.4$ K) at a temperature of $T=100$ mK: in this case, ideally, the quasiparticle density is phenomenally low $n_{\rm qp} \sim 10^{-5}$ $(\mu$m$)^{-3}$. However, invariably experiments have shown quasiparticle densities above what is predicted by \eref{eq:thqp} at the lowest temperatures (an example is shown in \fref{fig:qps}). These excess quasiparticles can be explained in two ways: They are either (i) created by external pair-breaking sources or (ii) there are sub-gap quasiparticle states not present in an ideal BCS-superconductor. According to our present understanding, the first option is the predominant one and the main external source is high frequency noise radiated from the environment. This is discussed in more detail in the next subsection. However, often the experimental observations can and have been interpreted by adopting the second interpretation.

The amount and dynamics of the excess (often referred to as non-equilibrium) quasiparticles has been under intense research lately, as they are a primary source of errors in almost all superconducting electronics. In addition to the anomaly at low temperatures, they are also produced when operating any superconducting device if there are dissipative elements in the circuit. Lately, these non-equilibrium quasiparticles have been considered in relation to qubits \cite{Palmer2007, Shaw2008, Martinis2009}, radiation detectors \cite{Barends2008, Visser2011}, single-electron turnstiles \cite{Saira2011} and NIS tunnel junctions \cite{Arutyunov2011}. All of these have confirmed the existence of excess quasiparticles at the lowest temperatures as well as the assumed dependence \eref{eq:thqp} of the thermal quasiparticles at higher temperatures.

\begin{figure}
\centering
\includegraphics{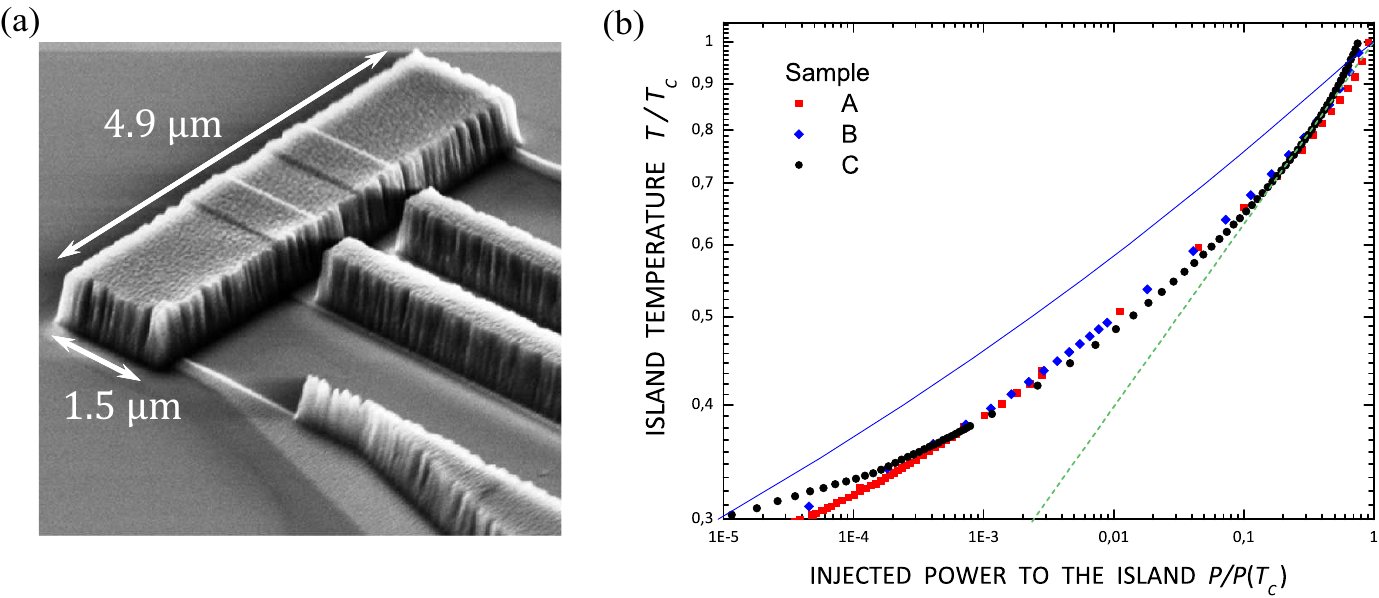}
\caption{Quasiparticle relaxation in Al, presented in \cite{Timofeev2009}.
\textbf{(a)} The experimental sample. Quasiparticle relaxation is probed in the thick central island, using two SIS junctions to inject the quasiparticles and two other junctions to probe the distribution function. The sample is fully aluminium.
\textbf{(b)} Measured $qp-p$ relaxation as a function of the electronic temperature. Dashed line shows the expected normal state relaxation and solid line is the result from full quasiclassical theory. Data from three different samples all lie between the normal state and the theory predictions.
}
\label{fig:AndreyBCS}
\end{figure}

Quasiparticle recombination is a process where two quasiparticles of opposite momenta ($\vec{k}$ and $\vec{-k}$ where $k \approx \Delta$) recombine to form a Cooper pair and emit a phonon with energy equal to $2\Delta$. The recombination rate was studied several decades ago \cite{Kaplan1976} but the associated heat flux from quasiparticles to the phonon system has been experimentally determined only very recently \cite{Timofeev2009}. At the limit where $T_p \ll T_{qp} \ll \Delta/k_B$, the heatflux can be calculated analytically from the quasiclassical theory to yield
\begin{equation}
P_{qp-p}^s \approx 0.98 \rme^{-\Delta/(k_BT_{qp})} P_{e-p}^n (T_{qp}, T_p),
\label{eq:approxPqp}
\end{equation}
where $T_{qp}$ is the quasiparticle temperature. However, experimentally the authors of \cite{Timofeev2009} found that the heat flux was larger than what was expected theoretically. The possible additional relaxation channels remain unclear at the moment (see \fref{fig:AndreyBCS} for the experimental set-up and results).

Now, working under the quasiequilibrium assumption, we can also write the heat diffusion equation in the superconducting case. The reduction in thermal conductivity at the superconducting state has been calculated theoretically soon after the BCS theory appeared \cite{Bardeen1959}. Assuming that the thermal conductivity is limited by impurities, in the superconducting state it can be written as $\kappa_s = \gamma(T) \kappa_n$, where the suppression ratio $\gamma(T)$ is given by
\begin{equation} \label{eq:gamma}
\gamma(T)=\frac{3}{2\pi^2}\int_{\Delta/k_BT}^\infty \frac{x^2}{\cosh^2(x/2)}dx \simeq \frac{6}{\pi^2}(\frac{\Delta}{k_BT})^2 e^{-\Delta/k_BT},
\end{equation}
where the approximation shown on the right again applies for $k_BT \ll \Delta$. Note that we assume everywhere that the superconducting gap $\Delta$ has the temperature dependence given by BCS theory. With these equations, a diffusion equation for the superconductor can be constructed just by inserting \eref{eq:approxPqp} and \eref{eq:gamma} to \eref{eq:diffN}
\begin{equation}
\nabla \cdot [-\kappa_s(x,T_{qp}) \nabla T_{qp}(x)] = \mathcal{P}_{qp-p}^s(x,T_{qp},T_p) + \mathcal{P}_{\rm ext}(x),
\label{eq:diffS}
\end{equation}
where again $\mathcal{P}_{\rm ext}$ is the power density from external sources and $\mathcal{P}_{qp-p}^s$ is $P_{qp-p}^s/\mathcal{V}$.

\begin{figure}
\centering
\includegraphics{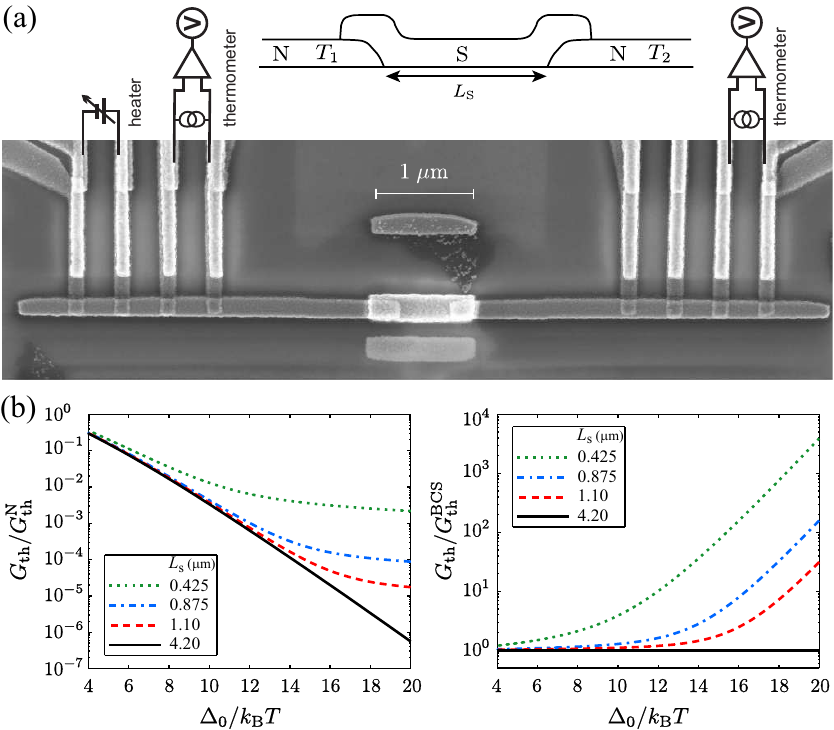}
\caption{Thermal conductance of a short superconducting (aluminium) wire between two normal metal (copper) leads, presented in \cite{Peltonen2010}. The thermal conductivity is enhanced by inverse proximity effect.
\textbf{(a)} SEM image of the sample and sketch of the measurement setup. Two Cu islands are connected via a short superconducting Al wire with transparent NS interfaces. Four S electrodes (top of the image) are connected to each of the two N islands through tunnel barriers for electronic thermometry and temperature control. Inset: Sketch of the side profile of the NSN structure, consisting of an S wire connected via overlap junctions to two N reservoirs. 
\textbf{(b)} Thermal conductivity of the S wire, normalised to the normal state conductivity (left) and to the BCS theory prediction (right). This data was extracted from fits to the measurement results. The four samples correspond to different lengths of the superconducting wire: 4.2, 1.1, 0.875 and 0.425 $\mu$m. The longest wire behaves as a BCS superconductor, whereas the heat conductivity of the shortest wire is many orders of magnitude larger. Two other samples fall in between these extreme cases.
}
\label{fig:proximity}
\end{figure}

The thermal conductivity of the superconductor can, however, be significantly modified if a normal metal is brought into contact with it \cite{Virtanen2007}. A transparent normal metal - superconductor contact will modify both the properties of the normal metal (proximity effect) and the superconductor (inverse proximity effect) close to the interface. In a superconductor, this will lead to effectively decreased superconducting gap as well as non-zero density of states inside the gap in the vicinity of the interface. This means that on short distances, quasiparticles with energy below the gap can also be transported to the superconductor (and are not Andreev reflected) which enhances the thermal conductivity. The length scale of this effect is roughly the superconducting coherence length $\xi$. In \cite{Peltonen2010}, the thermal conductivity of inverse proximised superconducting wire was studied in detail. A superconducting wire with length $L$ of the order of $\xi$ was placed in between two normal metal wires and a temperature difference was applied over it. The results of the thermal conductivity were in agreement with the theory predictions from quasiclassical theory. The longest wire ($L = 4.2$ $\mu$m) behaved almost as an ideal BCS superconductor, i.e. according to \eref{eq:gamma}, whereas the shortest wire ($L = 0.425$ $\mu$m) showed many orders of magnitude larger thermal conductance. These results are presented in \fref{fig:proximity}.

\subsection{Coupling of the electronic system to electromagnetic environment} \label{sec:photons}

Of more recent interest is the coupling of the electronic system via radiation \cite{Ojanen2007,Ojanen2008,Pascal2011}. This can manifest itself as heating/cooling of a conductor due to the presence of another resistive conductor at higher/lower temperature coupled to the one being monitored. The hot environment can also lead to photon assisted tunnelling. These are both naturally well known concepts but there has been some recent interest in these phenomena on the quantum level in mesoscopic structures, since they govern the ultimate heat-balance of the system.

\begin{figure}
\centering
\includegraphics{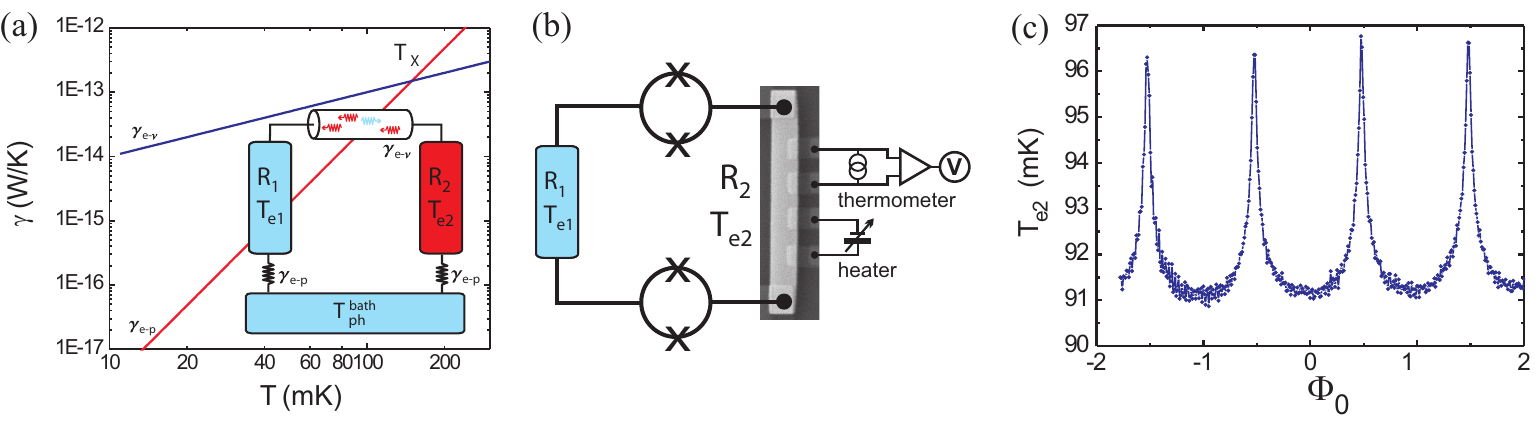}
\caption{\textbf{(a)} The inset of the figure shows the considered thermal model of the system: a resistor $R_1$ at electron temperature $T_{e1}$ is connected to a second resistor $R_2$ at higher temperature $T_{e2}$. The plot compares the thermal conduction via the electron phonon coupling ($\gamma_{e-p}$) to the phonon bath and heat conduction via thermal radiation ($\gamma_{e-\nu}$) for resistors with a volume of 4x0.25x0.03 $\mu$m$^3$. \textbf{(b)} Experimental setup: one resistor ($R_2$) has four tunnel probe contacts to allow thermometry and heat input. It is further connected to a second identical resistor without tunnel probes ($R_1$) via clean contacts to superconducting lines and SQUIDs. \textbf{(c)} Measured electron temperature of $R_2$ as a function of magnetic flux penetrating the SQUIDs.}
\label{fig:photon}
\end{figure}

Generally, radiative heat flux from a resistor $R_1$ at (electronic) temperature $T_1$ to resistor $R_2$ at (electronic) temperature $T_2$ via a transmission line with a total series impedance $Z_t(\omega)$ (see \fref{fig:photon} (b)) is described by \cite{Pendry1983}
\begin{eqnarray} \label{photoneq1}
P_{\nu}=&\int_{0}^{\infty}\frac{d\omega}{2\pi} \frac{4R_1R_2\hbar\omega}{|Z_t(\omega)|^2} \nonumber \\
& \times \left(\frac{1}{e^{\hbar\omega/k_BT_2}-1}-\frac{1}{e^{\hbar \omega/k_BT_1}-1}\right).
\end{eqnarray}
The Bose-Einstein distributions ${({e^{\hbar \omega/k_BT}-1})^{-1}}$ of the two resistors at the corresponding temperatures and the matching between them $r_0\equiv 4R_1R_2/|Z_t(\omega)|^2$ determine the total heat flux by the electromagnetic noise. Energy exchange with the environment via this photonic heat exchange can overcome the $e-p$ coupling below the crossover temperature 
$T_{\rm X} \sim [r_0\pi k_{\rm B}^2/(30\hbar \Sigma \Omega)]^{1/3}$ \cite{Schmidt2004}. 
For typical mesoscopic structures made of normal metals with a volume on the order of $\Omega \sim 10^{-20}$ m$^3$ and electron-phonon coupling strength $\Sigma \sim 10^9$ Wm$^{-3}$K$^{-5}$, one obtains moderately low values of $T_{\rm cr}\sim 150$ mK (see \fref{fig:photon} (a)). These parameters are within the range of experimental values for micron-scale refrigerators and have to be considered when describing the device operation. A strong coupling to a hot environment will degrade the cooler performance noticeably towards low temperatures. On the other hand, the total coupling strength can be minimized via the coupling $r_0$ between the impedance of the environment and the device as demonstrated experimentally in \cite{Meschke2006,Timofeev2009a}. We discuss the latter cooling experiment in more detail in \sref{sec:remote}. The experimental setup in \cite{Meschke2006} (see \fref{fig:photon} (b)) examines the influence of the radiative heat exchange as the coupling $r_0$ between the two coupled resistors is varied using SQUIDs. Their Josephson inductance $L_{\rm{J}}\simeq \hbar/(2eI_{\rm{C}})$ is influenced by the penetrating external magnetic flux ($\Phi$) through the SQUIDs as $I_{\rm_{C}}\propto\left|\cos(\Phi/\Phi_{\rm{0}})\right|$. $I_{\rm_{C}}$ is minimized at integer values of the flux quantum $\Phi_{\rm{0}}$, thereby maximising the Josephson inductance. This consequently minimizes the coupling $r_0$ between both resistors causing the measured temperature of $R_2$ to peak (see \fref{fig:photon} (c)).

It has been recently shown \cite{Pekola2010} that high-frequency noise radiated from higher temperatures $T_{\rm env}$ by some environment creates a leakage current with approximately linear bias voltage dependence in an NIS junction at low bath temperatures $T$, if $k_BT_{\rm env}\ge \Delta$. This leakage current is due to photon-assisted tunnelling events (see \sref{sec:brownian}), where high energy photons are absorbed during the tunnelling event and hence can facilitate lower energy electrons from N to tunnel above the gap to S. For a resistive environment with high cut-off frequency, this leakage current is exactly equivalent \cite{Pekola2010} to the one which one gets by assuming instead of the pure BCS density of states (DOS) $n_S(E) =|E|/\sqrt{E^2-\Delta^2}$, the so-called Dynes DOS with a lifetime broadening $\gamma$
\begin{eqnarray} \label{eq:DOSdynes}
n_S(E) =\Big{|}{\rm Re} \frac{E+i\gamma \Delta}{\sqrt{(E+\rmi\gamma\Delta)^2-\Delta^2}}\Big{|}.
\end{eqnarray}
Experimentally $\gamma$ can be extracted as the ratio of the measured zero bias conductance of the NIS junction and the asymptotic conductance at large voltages ($R_T^{-1}$).

The photon-assisted tunnelling events lead to $\gamma = 2\pi Rk_BT_\textrm{env}/(R_K\Delta)$, where $R$ is the effective resistance of the environment and $R_K = h/e^2$ the resistance quantum. In effect this linear behaviour amounts to viewing the subgap current of the NIS junction as if it were that of a fully normal (NIN) junction with tunnel resistance $R_T/\gamma$. Then the tunnelling rates at zero bias have a value $\Gamma_0= \frac{k_BT}{e^2R_T}\gamma$ and the power input is then roughly $\dot Q \simeq 2\Gamma_0\Delta$, where the factor $2$ appears because of tunnelling in two directions, each creating one quasiparticle. The radiated noise can then, at least partially, also explain the low-temperature anomaly of extra quasiparticles mentioned above. Experimental confirmation for this hypothesis has been added recently \cite{Barends2011,Paik2011,Saira2011}, as decreases in the amount of excessive quasiparticles at low temperatures have been seen to depend on the filtering and shielding of experimental set-ups.
\section{Cooling with NIS-junctions}

\subsection{Basics of NIS cooling and thermometry}

\begin{figure}
\centering
\includegraphics{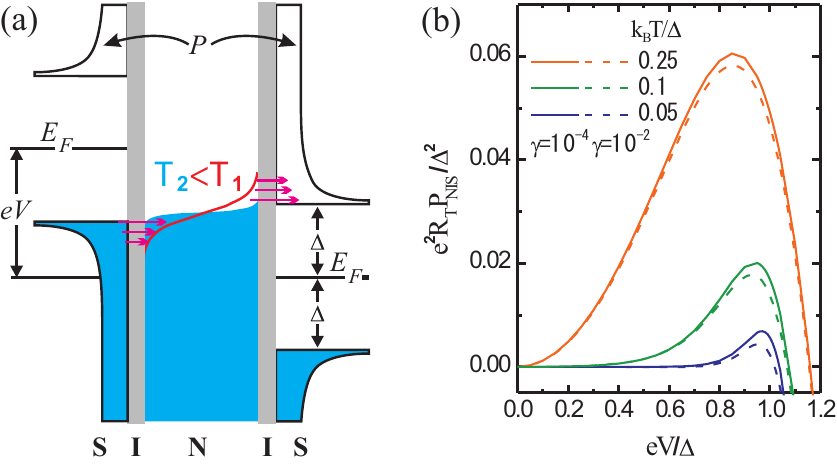}
\caption{Operational principle of a SINIS cooler.
\textbf{(a)} Energy diagram of a SINIS structure. When a proper bias voltage is applied, only hot electrons above the Fermi level can tunnel out from and only cold electrons below the Fermi level can tunnel into the normal metal. This leads to cooling of the electron system in the normal island.
\textbf{(b)} Normalised cooling power of the normal metal, as a function of applied voltage. Solid and dashed lines correspond to Dynes ($\gamma$) parameter values of $10^{-4}$ and $10^{-2}$, respectively. Different groups of lines correspond to different temperatures, from top to bottom $T=T_S=T_n$ is 0.25, 0.1 and 0.05 times $\Delta/k_B$. 
}
\label{fig:nis}
\end{figure}

The basics of cooling by NIS-junctions (N = normal metal, I = insulator, S = superconductor) have been discussed in several works, and many experiments have confirmed the predicted overall behaviour of this system (see \cite{Giazotto2006} and references therein). The phenomenon is based on the gap in the density of states (DOS) of a superconductor acting as an energy filter for the electrons. If proper bias voltage ($eV$ just below the gap energy) is applied over the junction, only the most energetic electrons can tunnel out from the normal metal (see \fref{fig:nis}). This leads to a decrease of the average energy of electrons in the N conductor, i.e. cooling. The opposite process, quasiparticle tunnelling from the superconductor to the normal metal has identical influence in terms of the energy content: it can be viewed as tunnelling of hole-like quasiparticles from N to S. The cooling power, i.e. the amount of heat extracted from a normal conductor per unit time in a NIS-junction biased at voltage $V$, is simply given by
\begin{eqnarray} \label{eq:nisP}
P_{\rm NIS} = \frac{1}{e^2R_T}\int \rmd E (E-eV)n_S(E)[f_N(E-eV)-f_S(E)],
\end{eqnarray}
where $R_T$ is the normal state resistance of the junction, $n_S(E)$ is the DOS in the superconductor (normalized by that of the corresponding normal metal), and $f_S(E),f_N(E)$ are the energy distributions of electrons in S and N, respectively. It should be noted that as \eref{eq:nisP} is symmetric in voltage, putting two NIS-junctions in series (forming a SINIS structure) doubles the cooling power of the normal metal. This is a different behaviour as compared to traditional Peltier cooling elements. Typically one assumes the basic BCS DOS $n_S(E) =|E|/\sqrt{E^2-\Delta^2}$. However, it is usually convenient, and sometimes also justified, to assume life-time type broadening of the BCS DOS so that it follows the so-called Dynes form with parameter $\gamma$ describing the sub-gap leakage (see \sref{sec:photons}).
The ideal behaviour of the cooler is achieved when the DOS is of pure BCS type, since in this case the electrons extracted from the normal metal can tunnel only to the states above the gap energy $\Delta$, leading to ideal energy filtering. Furthermore, to capture the ideal performance of the NIS-cooler as a refrigerator in a traditional sense, one assumes that the occupations of quasiparticles in S and N follow the Fermi-Dirac distribution, i.e. $f_{S,N}= 1/(1+e^{E/k_BT_{S,N}})$, where $T_S$ and $T_N$ are the (electronic) temperatures of the two conductors.

Consider an idealized cooler with pure BCS DOS and with well-defined temperatures $T_S$ and $T_N$, which may both differ from the bath temperature $T_0$. In this case, for temperatures well below the critical temperature of the superconductor $k_BT \ll k_B T_C\simeq \Delta /1.76$, one obtains analytical expressions for the optimal cooling power \cite{Anghel2001} (see also \cite{Mueller1997}).The cooling power maximizes at bias voltages $V = (\Delta-0.66k_BT)/e$ where it reaches 
\begin{equation} \label{eq:nisOpt}
P_{\rm opt}\simeq  \frac{\Delta^2}{e^2R_T}\Big[0.59(\frac{k_BT_N}{\Delta})^{3/2}- \sqrt{\frac{2\pi k_BT_S}{\Delta}}e^{-\Delta/k_BT_S}\Big].
\end{equation}
At the optimal bias point, the current through the cooler junction is (see below)
\begin{equation} \label{nis4}
I(V_{\rm opt})\simeq  0.48 \frac{\Delta}{eR_T}\sqrt{\frac{k_BT_N}{\Delta}}.
\end{equation}
An important figure of merit of the cooler is its coefficient of performance (``efficiency'') $\eta$, which we define as the cooling power at the optimum point divided by the power consumed in the voltage source, i.e.
\begin{equation} \label{nis5}
\eta = \frac{P_{\rm opt}}{I(V_{\rm opt})V}\simeq 0.7\frac{T}{T_C},
\end{equation}
where the last approximation applies again at $T\ll T_C$. Some of the characteristics of an ideal cooler have been shown in \fref{fig:nis}.

The current-voltage (I-V) characteristics of NIS-junctions are strongly non-linear and dependent on temperature, a fact that enables the use of these junctions also as thermometers. Consider the same kind of junction as mentioned above. At zero bias, the current is strongly suppressed due to the superconducting gap. As bias voltage is applied over the junction a (cooling) current starts flowing when the voltage is roughly $k_BT_N/e$ below the gap. This transition from the insulating to the conducting state depends heavily on the width of the Fermi distribution in the normal metal island, i.e. its temperature. With the same assumptions as above the current through a NIS junction can be written as
\begin{eqnarray} \label{eq:nisI}
I &=& \frac{1}{eR_T}\int \rmd E n_S(E)[f_N(E-eV)-f_S(E)] \nonumber \\
	&=& \frac{1}{2eR_T}\int \rmd E n_S(E)[f_N(E-eV)-f_N(E+eV)].
\end{eqnarray}
From the latter symmetrised form, it is clear that the I-V characteristics depend only on the temperature of the normal metal and are insensitive to the temperature of the superconductor (assuming constant superconducting gap). It should be noted that this insensitivity to the quasiparticle temperature of the superconductor does not apply to the cooling effect; no symmetrised form excluding $f_S$ can be derived from \eref{eq:nisP}.

As NIS thermometers are simple, easy to use and measure only the temperature of the normal metal, they are almost invariably used for thermometry in conjunction with NIS coolers. They do, however, have some drawbacks. Most notably, there is an inevitable power dissipation from the operation of the thermometer. This power can be made very small ($\sim$fW) but it can still have a notable influence on the temperature at the lowest temperatures the coolers are operated ($<$ 50 mK). In addition, NIS thermometers tend to lose sensitivity at the lowest temperatures, as the I-V characteristics are more influenced by the leakage currents through the junction ($\gamma$ parameter above). A proposed and demonstrated alternative to probe the lowest electronic temperatures is to measure proximity induced supercurrent in an SNS structure \cite{Dubos2001, Heikkila2002, Jiang2003, Courtois2008, Meschke2009} where the N island can be also connected to NIS cooler junctions. In this type of a structure, the measurement consists of sweeping current through the SNS system and measuring the critical current $I_c$ where the structure switches to the resistive state. The switching current depends strongly on the temperature of the N island and the measurement becomes increasingly sensitive towards low temperatures, as the supercurrent increases. Also, ideally the power dissipation is exactly zero before the switching happens, meaning that the thermometry has no self-heating effect. The approach also has some disadvantages, mainly that it is quite complicated and time consuming as in practice the measurement consists of making a switching histogram at each temperature point. In addition, as the exact dependency of the switching current from the temperature is a strong function of structure parameters, the extrapolation to lowest temperatures (where no calibration exists) is not straightforward.

Below we will give a review of the experimental and theoretical advances done with NIS cooling in recent years. For earlier history, see for example \cite{Giazotto2006}.

\subsection{Limitations on NIS cooler performance}

Generally the performance of NIS coolers has been below what would be expected from \eref{eq:nisP}, especially at temperatures below $\sim 150$ mK. Many possible reasons for this have been identified, here we review two most often raised effects, the excess density of quasiparticles in the superconducting leads and two-electron tunnelling processes.

\subsubsection{Effects of quasiparticle population on refrigeration} \label{sec:diffModels}

As mentioned above the efficiency of a NIS cooler (the ratio of the cooling power over the input power) is roughly $0.7 \: T/T_C$. At 0.3 K this corresponds to 15 \% (assuming $T_C$ of 1.3 K, common to thin Al films). Put another way, the power dissipated into the superconducting leads is (even in this ideal case and at rather high temperature) always an order of magnitude larger than the cooling power. In any practical cooler, this can be a significant power. As both the quasiparticle-phonon ($qp-p$) relaxation rate and the diffusion of quasiparticles are additionally exponentially suppressed in the superconductor (as described in \sref{sec:qps}), the dissipated power can create a high density of non-equilibrium quasiparticles on the superconducting side of the cooler, i.e. heat it up. This can have severe effects on the cooling power of NIS junctions.

\begin{figure}
\centering
\includegraphics{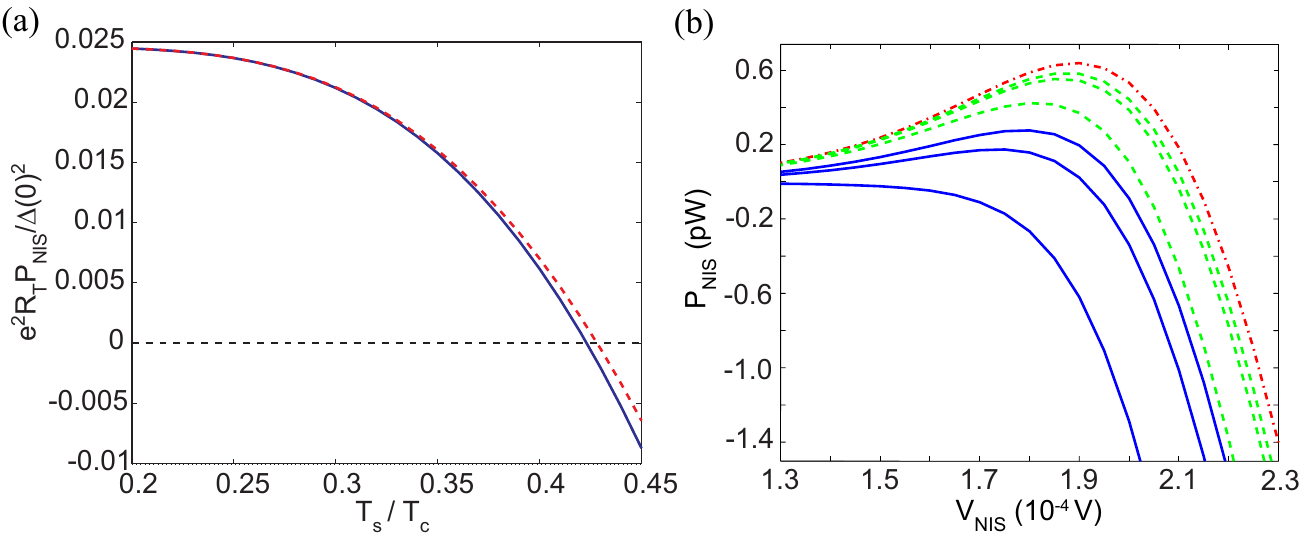}
\caption{Influence of superconductor temperature to NIS cooling. 
\textbf{(a)} $P_{\rm NIS}$ at the optimum cooling voltage as a function of $T_S$. $T_N$ is assumed to stay constant at 0.2 $T_C$. Solid line with BCS temperature dependent gap, dashed line with constant gap, both have $\Delta(0) = 200$ $\mu$eV and $\gamma = 10^{-4}$.
\textbf{(b)} $P_{\rm NIS}$ for junction with $R_T$ = 1 k$\Omega$ including the temperature rise in the superconductor calculated self-consistently with diffusion equations presented in Sec. 2. The top dash-dotted curve is the optimal ($T_s=T_\mathrm{bath}$) case. Solid lines correspond to different cross-sectional areas for the superconducting wire. From bottom 0.05, 0.15, and 0.25 $\mu$m$^2$. Dashed lines are calculated including a quasiparticle trap which starts from 1 $\mu$m distance from the junction through an oxide layer with resistivity 1 k$\Omega\mu$m$^2$. The cross-sectional areas are the same, thickness of Al is 50 nm. Other parameters: $T_{\rm bath} = T_N = 0.2$ K, $\Delta(0) = 200$ $\mu e$V and $\gamma = 10^{-4}$.
}
\label{fig:PvsTs}
\end{figure}

Assuming that non-equilibrium quasiparticles can be described with an effective temperature, the reduction of cooling power due to this overheating can be understood from \eref{eq:nisP}. In \fref{fig:PvsTs} (a), the cooling power of a NIS junction is presented as a function of the superconductor temperature at a constant normal metal temperature. The effect of rising $T_S$ is two-fold: it changes the distribution function $f_S$ (meaning there are more quasiparticles above the gap that can tunnel into the normal metal) and it reduces the $\Delta$ parameter in the DOS. Unlike for the electric current, for the heat flow the distribution function of superconductor plays a major role and the rising $T_S$ can destroy the cooling effect, even assuming a constant superconducting gap. The same effect is also visible in \eref{eq:nisOpt}, which has been derived assuming constant superconductor gap. Note that throughout the discussion we assume that the quasiparticle distribution still follows a Fermi distribution and is not in a true non-equilibrium.

In many instances, this overheating of the S has been modelled simply as a backflow parameter of heat, where a constant portion of the whole input power $IV$ is assumed to ``flow back'' to the normal metal and induce a parasitic heating power $P_{bf} = \beta IV$. Typically $\beta$ lies between 1 and 10 \%. Although this kind of a model has had some success in fitting the experimental data, it does not really address the mechanisms behind the backflow.

Recently, there has been considerable interest to model this effect more precisely, based on the diffusion equations presented in Sec. 2. Assuming a diffusion of quasiparticles away from the junction area as well as their relaxation (recombination), one can self-consistently calculate the effective temperature profile of the superconductor and hence the cooling power of the junction. In \fref{fig:PvsTs} (b), we present a calculation of the cooling power using the effective temperature model presented in Sec. 2.

To make some simple estimates with a diffusion model, let us consider a 1D temperature profile and make the assumptions $T_p \ll T_{qp} \ll \Delta/k_B$. Then we can write the diffusion equation for a superconductor in an analytically solvable form \cite{Timofeev2009a} (we neglect the $ P_{\rm ext}$ term and the prefactor of the order of unity in \eref{eq:approxPqp})
\begin{equation} \label{eq:SdiffApprox}
\frac{\partial}{\partial x}[\frac{6}{\pi^2}(\frac{\Delta}{k_BT})^2 \rme^{-\Delta/(k_BT)} \kappa_n \frac{\partial T}{\partial x}]
 = \rme^{-\Delta/(k_BT)} \Sigma (T^5-T_p^5),
\end{equation}
where we have written $T = T_{qp}$ for clarity. Linearising \eref{eq:SdiffApprox} for small temperature differences $\delta T(x)= T(x)-T_p$, one obtains a simple expression for the temperature profile in a uniform one-dimensional wire. For a wire extending to positive $x$, we have then $\delta T(x) = \delta T(0) e^{-x/\ell}$, where $\delta T(0)$ is determined by the heat input at the end of the wire and the relaxation length is given by
\begin{eqnarray} \label{eq:relax}
\ell = \frac{\Delta}{\pi k_B}\sqrt{\frac {6 \mathcal{L} \sigma}{5\Sigma}}T_p^{-5/2}.
\end{eqnarray}
Putting the parameters of aluminium in \eref{eq:relax}, we find that $\ell \simeq (50$ $\mu$m$\cdot$K$^{5/2})/T^{5/2}$. This means that at typical sub-kelvin temperatures, the quasiparticle distribution relaxes over millimetre distances. The magnitude of the temperature rise can be obtained in the same linearised approximation by employing the boundary condition
$P = -\kappa_s A \frac{d\delta T(x)}{dx}|_{x=0}$,
where $A$ is the cross-sectional area of the wire. Inverting this for the temperature rise for a given heat input, we find
\begin{eqnarray} \label{ne7}
\delta T(0) = \frac{l}{\kappa_s} \frac{P}{A} = \frac{\pi k_B}{\Delta \sqrt{30 \mathcal{L} \sigma \Sigma}} \rme^{\Delta/(k_BT)} T^{-3/2} \frac{P}{A}.
\end{eqnarray}
Inserting numbers for a $A =100$ nm $\times$ 100 nm wire at $T= 200$ mK, yields $\delta T(0) \simeq 20 P/A \simeq (2\cdot 10^{15}$ K$\cdot$W$^{-1}$)$P$. This means that in order to keep $\delta T(0) \ll T$, one needs to have $P \ll 10^{-16}$ W, i.e. a very small power input indeed. Assuming each quasiparticle brings energy $\Delta$ to the superconductor, this implies a tunnelling rate of $\Gamma = P/\Delta \ll 3\times10^6$ s$^{-1}$. This corresponds to bias current of only 0.5 pA. This example demonstrates that a bare superconducting wire is driven out of equilibrium even with a very small current injection.

It is instructive to compare this approach (using effective temperature), with the results one gets using just the quasiparticle density. In relation to micron scaled coolers, this approach was first used in \cite{Ullom1998} and has been expanded in \cite{Rajauria2009, Oneill2011}. In this case the diffusion equation in 1D is
\begin{equation} \label{eq:Ullom}
D_s \frac{\partial^2 n(x)}{\partial x^2} = \Gamma_{qp-p} + \Gamma_{\rm ext},
\end{equation}
where $\Gamma_{qp-p} + \Gamma_{\rm ext}$ are now the relaxation (scattering) rates to phonons and external environment respectively and can be converted to power by multiplying with the energy exchanged in each scattering event. $D_s$ can be related to the normal state diffusion constant $D_n$ by $D_s = \sqrt{1-(\Delta/E)^2} D_n$, where $E$ is the energy of the quasiparticle and $D_n$ is related to normal state heat conductivity through $\kappa_n = \mathcal{L} D_n N_F e^2 T$. In this way, one does not need to make the assumption of quasi-equilibrium but one now needs to consider explicitly the energies of the quasiparticles. In practice, some assumptions are needed. One can either assume that the quasiparticles follow a thermal distribution and, in fact, it is straightforward to show that in that case the left-hand-side (LHS) of \eref{eq:Ullom} is exactly equivalent to LHS of \eref{eq:SdiffApprox} (the connection between $n$ and $T$ is from \eref{eq:thqp}). The other option (adapted in \cite{Ullom1998} and \cite{Rajauria2009}) is to replace the $E$ in \eref{eq:Ullom} with the average energy of quasiparticles in the sample $\langle E \rangle$. This makes the $D_s$ independent of $x$ coordinate and essentially is an approximation for small temperature differences where the spatial dependence of the diffusion constant can be neglected.

From \fref{fig:PvsTs} (b) and equations \eref{eq:relax} and \eref{ne7}, it is clear that for the design of efficient NIS coolers, it is critical to consider also the quasiparticle thermalisation of the superconducting leads. Fortunately, the situation is usually not as bad as the solid lines in \fref{fig:PvsTs} (b) would seem to suggest, as the effects of so-called quasiparticle traps were neglected. These are usually normal metal (or lower gap superconductor) films which are in contact with the superconductor and act as heat sinks where the quasiparticles can be absorbed. The effect is based on the fact that (as described in \sref{sec:qps}) the normal metal has exponentially stronger electronic heat diffusion and electron-phonon coupling than the superconductor and hence the excess heat is quickly absorbed to the bath. However, a normal metal island very close to the junction can severely decrease the performance of the cooler as the energy gap of the superconductor is smeared due to the proximity effect. Hence, optimising the distance of the traps is difficult \cite{Pekola2000, Court2008}.

A safe way of introducing a trap with moderate improvement in quasiparticle relaxation rate comes for free in junctions fabricated by shadow angle deposition (if some care is taken in designing the leads in the vicinity of the junctions). This fabrication procedure produces first the superconducting (e.g. aluminium) lead, which is subsequently oxidized, and thereafter a metal layer (e.g. copper) is deposited at another angle, forming the NIS junction. Such an overlap structure can be made in the same process outside the junctions to cover partially the superconducting leads by the normal metal via the oxide barrier. The mechanism of the quasiparticle thermalization in this structure is via tunnelling of hot quasiparticles into the normal metal. The dashed lines in \fref{fig:PvsTs} (b) show the same calculation as the solid lines but now including also a quasiparticle trap. The improvement is considerable, even though the trap is assumed to be behind a relatively thick (1 k$\Omega\mu$m$^2$) oxide barrier.

By similar arguments used in obtaining \eref{eq:relax}, we can obtain a thermal relaxation length with the trap,
\begin{equation} \label{ne8}
\ell = \left( \frac{2\sqrt{2} d \rho_T \sigma}{\sqrt{\pi}} \right)^{1/2} \left( \frac{k_B T}{\Delta} \right)^{1/4}.
\end{equation}
Here $d$ is the thickness of the superconducting lead and $\rho_T$ is the specific resistivity of the trap barrier. For a relatively resistive barrier $\rho_T =1$ k$\Omega\mu$m$^2$ with  $d = 30\ \mathrm{nm}$ and $T = 200\ \mathrm{mK}$, we obtain $l \approx 20\ \mathrm{\mu m}$, which is about two orders of magnitude shorter than in a bare superconducting wire.

In \cite{Rajauria2009}, the diffusion model using quasiparticle density was extended to include the relaxation due to quasiparticle traps and the results seem to agree with experiments \cite{Rajauria2011}. Recent paper \cite{Oneill2011} considers both the quasiparticle density in the superconductor, the temperature profile of the normal metal trap and the possible ``athermal phonons'', i.e. very hot phonons emitted by quasiparticle recombination.

\subsubsection{Influence of Andreev current on refrigeration}

All the equations given above for the current and heatflow through a NIS interface have been derived assuming single-electron tunnelling events through the barrier. This approximation is generally valid for NIS junctions as proven by the good agreement between experiments and \eref{eq:nisI}. However, when going to very high transparencies of the junction (in order to maximize the cooling power), a second order process, called Andreev current, starts to be a significant transport mechanism. The Andreev current is essentially a process where a Cooper pair in the superconductor is transported into two quasiparticle excitations in the normal metal or vice versa. It can dominate over the single-particle current at voltages $\ll \Delta/e$.

\begin{figure}
\centering
\includegraphics{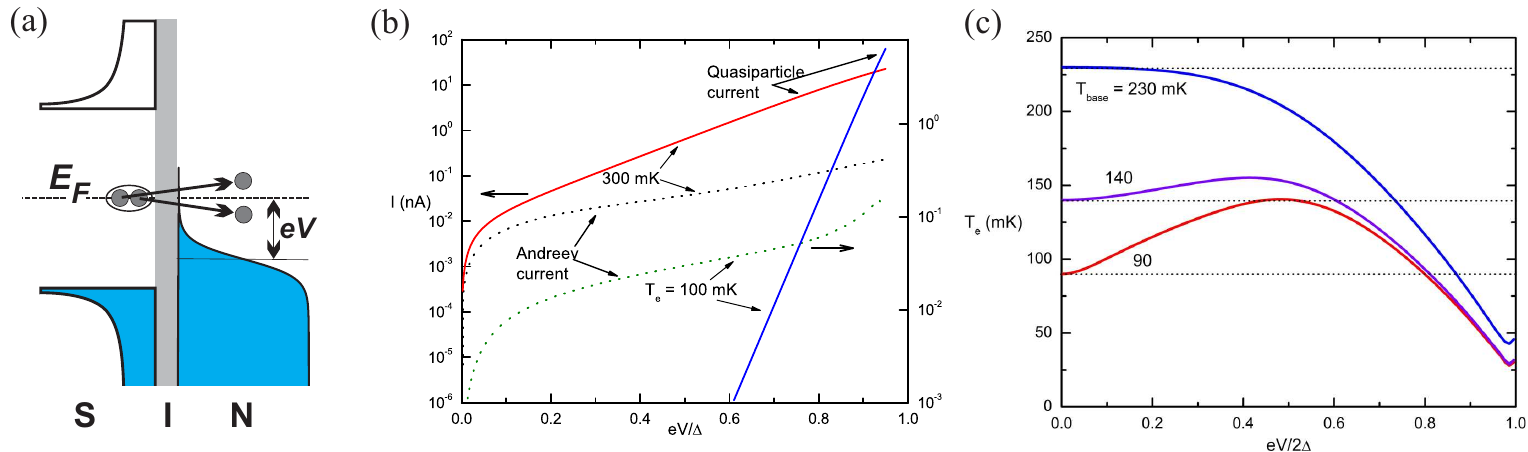}
\caption{Andreev current.
\textbf{(a)} Simplified energy diagram of Andreev process. A Cooper pair in the superconductor is transported through the interface and creates two quasiparticle excitations into the normal metal with average energy $eV$.
\textbf{(b)} Calculated current-voltage characteristic of the NIS junction with $R_T = 2$k$\Omega$ adapted from \cite{SukumarThesis}. The complete lines show the contribution due to quasiparticle current and the dotted lines show the phase-coherent Andreev current for a diffusive junction. Andreev current is significant only at low bias voltages and low temperatures.
\textbf{(c)} Influence of Andreev current on the normal metal temperature. The curves show the calculated electronic temperature as a function of the voltage, including the heating by the Andreev current, which is clearly visible at voltage $V < \Delta/e$. Parameters are derived from the fit to the experimental result.
\textit{(c) is a reprinted figure with permission from \cite{Rajauria2008}. Copyright 2008 by the American Physical Society.}
}
\label{fig:rajauria2008}
\end{figure}

The Andreev current through a normal-superconductor interface has been studied over several decades. The heat current through a NIS junction was studied theoretically by Bardas and Averin \cite{Bardas1995}, who showed that Andreev current (unlike the regular single-electron current) leads to dissipation in the N electrode at all bias voltages. This can be understood from a simple energy diagram picture, depicted in \Fref{fig:rajauria2008} (a). A Cooper pair on the Fermi level in the superconductor creates two excitations in the normal conductor with average energy is $eV$, where $V$ is the bias voltage across the junction. Therefore all the power $P_{\rm AR}$ dissipates in N and it equals simply $P_{\rm AR} = I_{\rm AR}V$, where $I_{\rm AR}$ is the electrical current due to the Andreev process. 

The magnitude of the Andreev current depends on several parameters of the tunnel junction and its electrodes. For small junctions at the ballistic limit (meaning that the dimensions of the junctions are smaller than the mean free path of electrons in the normal metal), it is proportional to voltage such that $I_{\rm AR}=(16\mathcal N R_T^2)^{-1} R_KV$, where $\mathcal N$ is the number of conduction channels, and $R_K=h/e^2$ is the quantum resistance \cite{Averin2008}. This ballistic description gives typically very small values for Andreev current with the transparencies common in NIS junctions. However, for larger diffusive junctions typical of NIS coolers, the Andreev current is not given by this simple expression. This is basically because disorder in the metals leads to quasiparticle confinement near the interface and they can experience multiple reflections before escaping the junction area. This can lead to orders of magnitude higher values of the Andreev current because of constructive interference between the consequent tunnelling amplitudes. Note that although the same confinement is also present in the single particle case, it gives no enhancement to current as there is no interference between the tunnelling amplitudes. The diffusive case can be analysed theoretically \cite{Hekking1993,Hekking1994,Vasenko2010} and the results seem to agree with experiments \cite{Rajauria2008, Lowell2011}.

Andreev current can have a significant influence on the cooling power and hence on the temperature of the N electrode at low bias voltages $|eV|\ll \Delta$ (see \fref{fig:rajauria2008} (b) and (c)). In practice, the N island of a SINIS cooler heats up at these intermediate voltages. Yet the contribution of Andreev current becomes less significant at voltages close to the optimal one at $\sim 2\Delta/e$, and, luckily enough, the achieved temperature reduction is generally almost unaffected by the Andreev effect.

\subsection{Coulomb blockaded NIS cooler}

In the basic description of a tunnel junction refrigerator, the Coulomb effects are neglected. However, as the size of the cooled normal metal volume is decreased and the tunnel barriers are made more opaque, the energy cost of extracting or putting electrons in the normal metal can become considerable. Although this is not a desirable effect for a basic cooler, it does bring up the possibility of controlling the heatflow with a gate, i.e. a heat transistor in analogue to a single-electron transistor (SET) for charge transport. 

The energy cost of adding or extracting one electron from a normal metal island, connected to metallic leads through tunnel barriers, is determined by the charging energy $E_C = e^2/(2C_\Sigma)$, where $e$ is the electron charge and $C_\Sigma$ the total capacitance of the island. This total capacitance includes both the stray capacitance of the island (determined by the size of the island) and the capacitance of the tunnel junctions (determined by the size of the junctions). In the regime where $E_C \gg k_BT$, there can be no electron tunnelling events to or from the island, unless the bias voltage over the island overcomes this charging energy cost. This is the effect of Coulomb blockade.

When combined with superconducting leads, a Coulomb blockaded SINIS device is created. One then has two energy barriers to overcome in order to be able to push current through the device: the superconducting gap and the charging energy. There is, however, one crucial difference between these two: the charging energy level of the island can be adjusted with an external gate. In this way another control parameter is added to the SINIS device and one can now tune the electron current and, hence, the heatflow at a certain bias point by adjusting the gate voltage. Theoretically, the maximum on/off ratio of heatflow at the optimum cooling point reachable this way can be written at the limit $E_C \gg k_BT$ as \cite{Saira2007}
\begin{equation}
P_\textrm{open}/P_\textrm{closed} \simeq 0.45\frac{k_BT}{E_C} \rme^\frac{E_C}{k_BT},
\end{equation}
which for example for $E_C/k_BT = 10$ gives a ratio of 990. Note, however, that even at the gate open position, the maximum cooling power of a SINIS structure is only one half of the corresponding case where Coulomb blockade plays no role, as heatflow through the other junction is effectively blocked.

\begin{figure}
\centering
\includegraphics{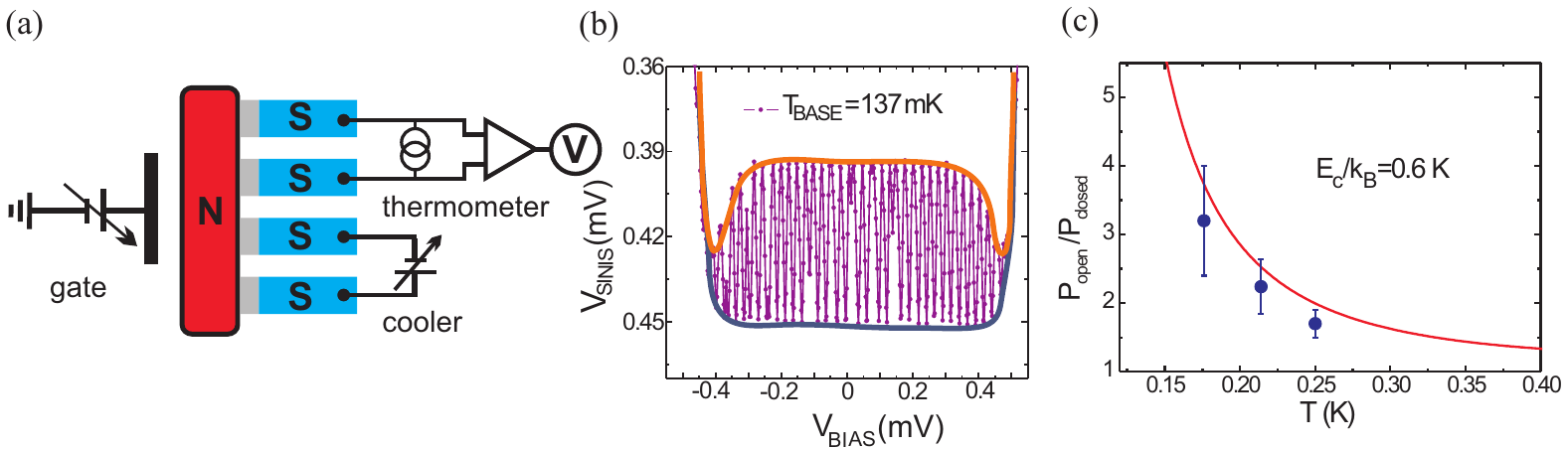}
\caption{Heat transistor.
\textbf{(a)} A schematic of the measurement setup.
\textbf{(b)} SINIS thermometer readout (dots connected with lines) as a function of applied cooler voltage under constant gate sweep condition. Gate open (orange) and gate closed (blue) lines are connected with lines. A cooling peak is present only for the gate open case. Note that the SINIS probe voltage depends both on the gate charge and the temperature, for details see \cite{Saira2007}.
\textbf{(c)} Theoretical (line) and experimental (dots) values for the on/off ratio $P_\textrm{open}/P_\textrm{closed}$ for the cooler used in \cite{Saira2007}. The charging energy of this particular structure was relatively low, leading to modest on/off ratios.
}
\label{fig:heatTrans}
\end{figure}

The heat transistor effect was experimentally demonstrated in \cite{Saira2007}, where an on/off ratio of over three was demonstrated (see \fref{fig:heatTrans}). In the demonstration, Coulomb blockaded SINIS structures were used for both thermometry and cooling. The correspondence between simulations based on Coulomb-blockaded single-electron tunnelling (orthodox theory) and measurement was compelling, proving that SINIS structures can be used for thermometry also in this regime, although this requires a careful modelling of the charge distribution on the island.

Another possibility arising in the charging energy dominated regime, again analogously with electric current, is the pumping of heat. In recent years, there has been a lot of interest in using superconductor-normal metal SETs as a current standard by pumping single electrons through it thus connecting frequency to current by the relation $I = ef$ \cite{Pekola2008}, where $f$ is the pumping frequency. The hybrid SET has a definite advantage as compared to the normal metal SET: during one full pumping cycle between two stable charge states, one does not need to go through any point where current would be freely transferred, unlike in the normal case. This is because the parameters can be chosen so that the current is blocked by the superconducting gap at the intermediate gate voltages. This allows one to construct an accurate current turnstile using only one island.

\begin{figure}
\centering
\includegraphics{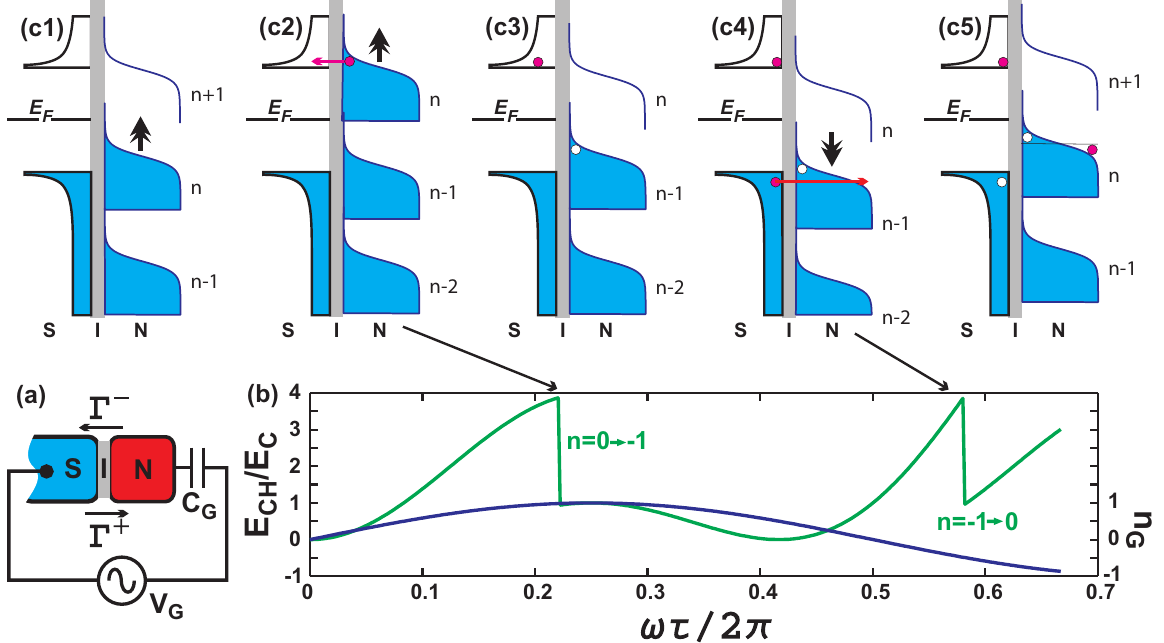}
\caption{Single-electron refrigerator. 
\textbf{(a)} Schematic principle: a small normal metal island is separated from one superconducting lead by a tunnel junction. The electrostatic potential of the island is controlled by a gate.
\textbf{(b)} A sinusoidally varied gate signal (right axis) with a suitable amplitude can allow tunnelling events to occur: from above the Fermi level in the normal metal to above the gap in the superconductor [\textbf{(c1)} to \textbf{(c3)}] and from below the gap to below the Fermi level \textbf{(c4)} and \textbf{(c5)}. The total charging energy ($E_{\rm CH}$) of the device [\textbf{(b)}, left axis] is reduced by the tunnelling events. On average, the energy taken by each of these tunnelling events from the normal metal is $k_BT$ and hence a driving signal with frequency $f$ will lead to cooling power $fk_BT$.}
\label{fig:SER}
\end{figure}

The same principle can also be applied to heatflow. The idea was presented in \cite{Pekola2007} (before the current standard proposition) and experimentally demonstrated in \cite{Kafanov2009}. If a sinusoidally varying gate signal with proper amplitude is applied to a Coulomb-blockaded SINIS structure, a sequential tunnelling of single-electrons with certain energy can be achieved. Each tunnelling will take (on average) an energy $k_BT$ from the island, producing a total cooling power of $k_BTf$.  A schematic of the process is presented in \fref{fig:SER}. Although the cooling power achievable this way is below the constant bias case, this kind of cooler has the obvious advantage that no bias voltage is in principle needed over the island and there is no need for a galvanic connection between the leads and the island. Also, it is an example of cyclical conversion of work into cooling, illustrating basic thermodynamics at the nanoscale.

\subsection{Brownian refrigerator} \label{sec:brownian}

Above we have discussed how a NIS junction can be used for cooling by either DC biasing it or by AC driving it (in the Coulomb blockaded case) through a capacitive coupling. However, a third option exists. If one can couple the NIS junction to a hot environment, the voltage fluctuations of the environment can ``drive'' the cooler. This leads to a somewhat counterintuitive phenomenon (which also at the first sight seems to defy the laws of thermodynamics) where a coupling to a hot environment cools the normal metal.

This kind of a Brownian refrigerator was proposed in \cite{Pekola2007a} (see also \cite{VandenBroeck2006}) and further extended in \cite{Peltonen2011}. There the authors considered a case where one couples a NIS junction to a hot resistor with superconducting leads providing ideally only photonic energy exchange between the resistor and the NIS structure. The hot photons emitted by the resistor can then ``kick'' the electrons to overcome the energy barrier needed for them to tunnel to the superconductor (no voltage is applied over the junction). If the temperature of the resistor is properly set, then preferentially only the high energy electrons are removed from the normal metal and this again leads to cooling. 

Quantitatively, one way to model this effect is the so-called $P(E)$ theory. Environmentally-assisted tunneling (or photon assisted tunnelling), where photons are exchanged with the electromagnetic environment during the tunneling event, has been studied extensively in relation to tunnel junctions \cite{Ingold1992}. In the case considered here (un-biased NIS-junction), the heatflow from the normal metal can be shown to follow
\begin{eqnarray}
P_n =& \frac{2}{e^2R_T} \int_{-\infty}^{\infty} \int_{-\infty}^{\infty}  \rmd E \rmd E' n_S(E') E \nonumber \\
& \times f_N(E)[1-f_S(E')] P(E-E'),
\label{eq:Pbrown}
\end{eqnarray}
where $P(E)$ is essentially the probability density of the environment to emit a photon with energy $E$. If the environment is highly resistive, so that $R_{env} \gg R_K$ where $R_K$ is the quantum of resistance, the $P(E)$ function is a gaussian centered around $E_C$ with a width that is temperature dependent. On the other limit where $R_{env} \ll R_K$, $P(E)$ is a $\delta$ peak at $E$ and \eref{eq:nisP} is recovered. 

\begin{figure}
\centering
\includegraphics{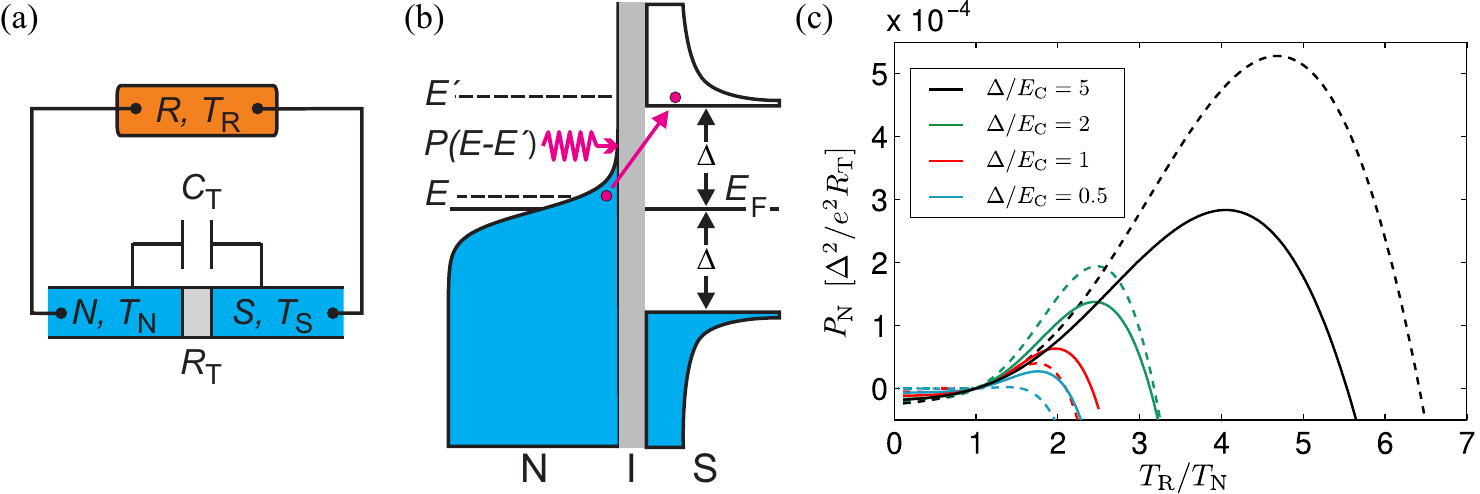}
\caption{Brownian refrigerator presented in \cite{Pekola2007a, Peltonen2011}.
\textbf{(a)} Basic circuit diagram. Hot resistor at temperature $T_R$ is connected to a NIS junction with junction resistance $R_T$, and provides energy allowing it to work as a cooler.
\textbf{(b)} Energy diagram of photon assisted tunnelling. Energy $E'-E$ is provided by the environment (the hot resistor) allowing an electron to tunnel to the superconductor from the normal metal.
\textbf{(c)} Calculated cooling power of the Brownian refrigerator $P_n$ as a function of the resistor (environment) temperature, calculated with different values of the charging energy $E_C$. Solid lines correspond to junction resistance $R_T = 0.5 R_K$ and dashed lines to $R_T = 10 R_K$. 
}
\label{fig:BR}
\end{figure}

The calculated heat flow out of the normal metal as a function of the environment temperature is presented in \fref{fig:BR} (c). The cooling power is maximised when the temperature of the resistor is around $T_N \Delta/E_C$. It should also be noted that when the resistor is cooler than the normal metal, there is a net heatflow from the superconductor to the normal metal, leading to cooling of the superconductor. This is a regime which is unattainable in a voltage biased NIS junction.
 
The Brownian refrigerator is still to be experimentally demonstrated. In actual experiment, there are some complications not included in the ideal treatment above. First, in practice the hot resistor must be heated with electric current, which then has to be prevented from flowing through the NIS junction. Second, finite frequency impedance of the circuit dictates that the heated resistor has to be quite close to the cooler junction, making heatflow through the phonon system a considerable concern. And third, if the resistance of the normal metal island is not totally negligible compared to the resistance of the cooling junction, there will be a parasitic heating power from direct photonic heating from the hot resistor. Nevertheless, taking all these effects into account, it was concluded in \cite{Peltonen2011} that the effect should still be experimentally detectable with realistic parameters.

\subsection{Remote cooling} \label{sec:remote}

\begin{figure}
\centering
\includegraphics{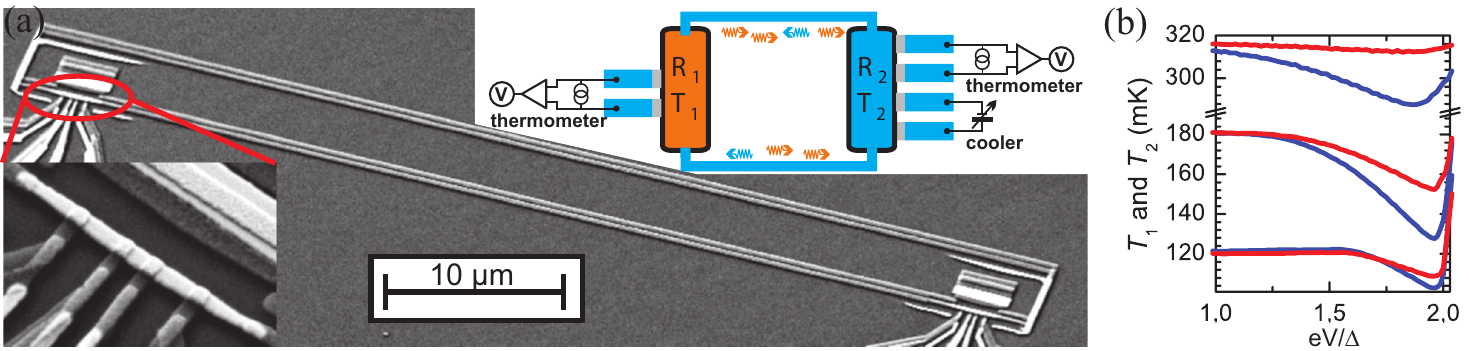}
\caption{Remote cooling, presented in \cite{Timofeev2009a}.
\textbf{(a)} A SEM picture of the sample and right inset showing schematic of the operating principle. Two normal metal islands are connected in a long superconducting loop. Both islands' electronic temperature is probed with NIS thermometers and the other island is also cooled with NIS junctions. Coupling through the transmission line couples the two temperatures at low bath temperatures.  
\textbf{(b)} Experimental results. Lower lines are the temperature of the directly cooled island and the upper lines are the temperature of the other island, located at 50 $\mu$m distance.
}
\label{fig:remote}
\end{figure}

Interestingly, the concept of radiative heat exchange introduced in \sref{sec:photons} allows the spatial separation of the micron-scale cooler and the cooled object itself when both objects are coupled together in a matched circuit \cite{Timofeev2009a}. The scheme in \fref{fig:remote} depicts the device concept: the actively cooled metal acts as a cold environment for the device, the latter is cooled via a transmission line. Thermometry of both the cooler island and the remotely connected island is done with standard SINIS thermometry. Above about 300 mK, heat is mainly transported between both islands by quasiparticles as the quasiparticle population is not frozen out yet. Superconducting aluminium employed in the experiment as transmission lines provides sufficient thermal isolation, from heat transported by quasiparticles, only at the lowest temperatures. Nevertheless, towards lower temperatures, when in addition the electron-phonon coupling diminishes, the photonic heat transport becomes dominant and couples both islands effectively. This results in the temperature drop of the remotely cooled island reaching about 60 $\%$ of the directly cooled island. Moreover, a complete galvanic isolation of cooler and the cooled device would be achievable if the structures are coupled capacitively or inductively to realize cooling at a distance.

\subsection{Effect of magnetic field to NIS cooling} \label{sec:magn}

When a magnetic field is applied over a superconductor, it will create surface currents that will cancel the field completely within a penetration depth from the surface, this is the well-known Meissner effect. In a type I superconductor, such as bulk Al, there exists a single well defined critical field above which the superconductivity is totally suppressed and below which the Meissner effect prevents magnetic field from entering the bulk of the superconductor. The situation changes, however, when the dimensions of the metal film become comparable to the penetration depth. In this thin wire form all superconductors display type II behaviour, having two critical fields. At the lower critical field $B_{c1}$, magnetic vortices start to penetrate the material, creating areas where the superconducting energy gap is locally suppressed but the overall superconducting behaviour is retained. The superconductivity is suppressed only at the higher critical field $B_{c2}$. In addition, the lower critical field is not primarily determined by the material but by the geometry of the wire. It has a universal characteristic value $B_{c1} \sim \Phi_0/W^2$, where $\Phi_0 = h/2e$ is the flux quantum and $W$ is the width of the wire (assuming a wire with thickness $\ll W$ and a magnetic field perpendicular to the wire) \cite{Stan2004}.

\begin{figure}
\centering
\includegraphics{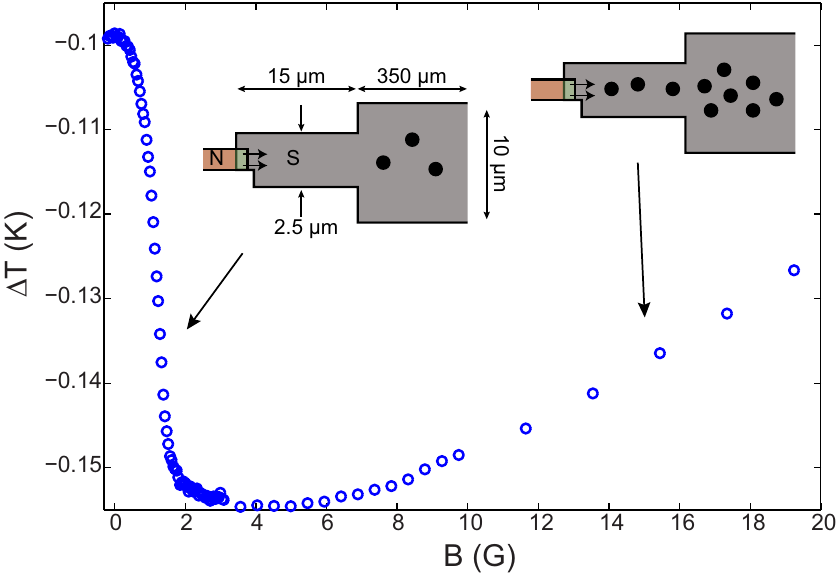}
\caption{Magnetic field effect to NIS cooling, presented in \cite{Peltonen2011a}, showing the maximum cooling $\Delta T$, starting from bath temperature of 285 mK, as a function of applied perpendicular magnetic field. The cooling effect is seen to be enhanced by over a factor of 1.5 by applying a small magnetic field. This is explained as magnetic vortices acting as quasiparticle traps and preventing the overheating of the superconducting lead. In larger fields vortices are created at the junction area, hence degrading the cooling effect.
}
\label{fig:Bfield}
\end{figure}

It has been shown that the creation of magnetic vortices leads to faster quasiparticle relaxation \cite{Ullom1998a}. The vortices act as quasiparticle traps: the areas with locally suppressed gap have also correspondingly higher $e-p$ coupling and can absorb the ``hot'' quasiparticles. Combined with the feature that the vortices will first be introduced into the widest parts of the superconductor, this allows one to design the superconducting leads of a NIS cooler so that the thermalisation of the lead will be optimised in a small magnetic field. This effect was recently reported in \cite{Peltonen2011a}, where a very significant improvement of NIS cooler performance was seen in small magnetic fields, if the lead geometry was designed so that it is narrower at the junction area than elsewhere (see \fref{fig:Bfield}). In the opposite case, vortices are first created at the junction area and will deteriorate the cooler performance.

\subsection{Cooling phonons with NIS coolers} \label{sec:membr}

A concern of NIS-coolers in practical applications has been the fact that they cool directly the electron system. In many cases, it would be desirable that the cooler structure itself is electronically isolated from the sample to be cooled. In order to achieve this, one has to somehow thermally couple the sample to the cooled normal metal volume without electrically coupling it. In practice, this means coupling through a phonon system. However, in order to cool a phonon system with NIS-cooler, one has to make the coupling from the phonon system to the environment smaller than the electron-phonon coupling in the normal metal. As the $e-p$ heat current decreases as $T^5$ at low temperatures, this is a very challenging condition.

The most straightforward way to achieve this isolation from the environment is to have the phonon system as a micromachined membrane, on top of which the samples to be cooled are fabricated. This membrane can then be cooled with so-called cold fingers, normal metal leads extending from a NIS-cooler to the membrane. The junctions itself need to be located on the bulk, in order not to let the heat dissipated to the superconductor to couple to the cooled volume.

This kind of a membrane cooler would be of considerable interest in many applications of superconducting electronics, ranging from quantum information technology to space borne radiation detectors. In principle, all of the community utilizing aluminium as a superconductor are facing a technological challenge in providing below 0.1 K temperatures where the superconducting properties of Al are optimized. Current solutions, mainly adiabatic demagnetization refrigerators and dilution refrigerators, are complicated to use and, more importantly for space applications, heavy. It would be enormously advantageous to replace these refrigerators with a simple $^3$He refrigerator, or even better, a pumped $^4$He bath, combined with a NIS membrane cooler. The first applications to benefit would be the ones where the fabrication onto a membrane is straightforward. This group includes especially radiation detectors which are already often fabricated on a membrane.

\begin{figure}
\centering
\includegraphics[width=0.5\textwidth]{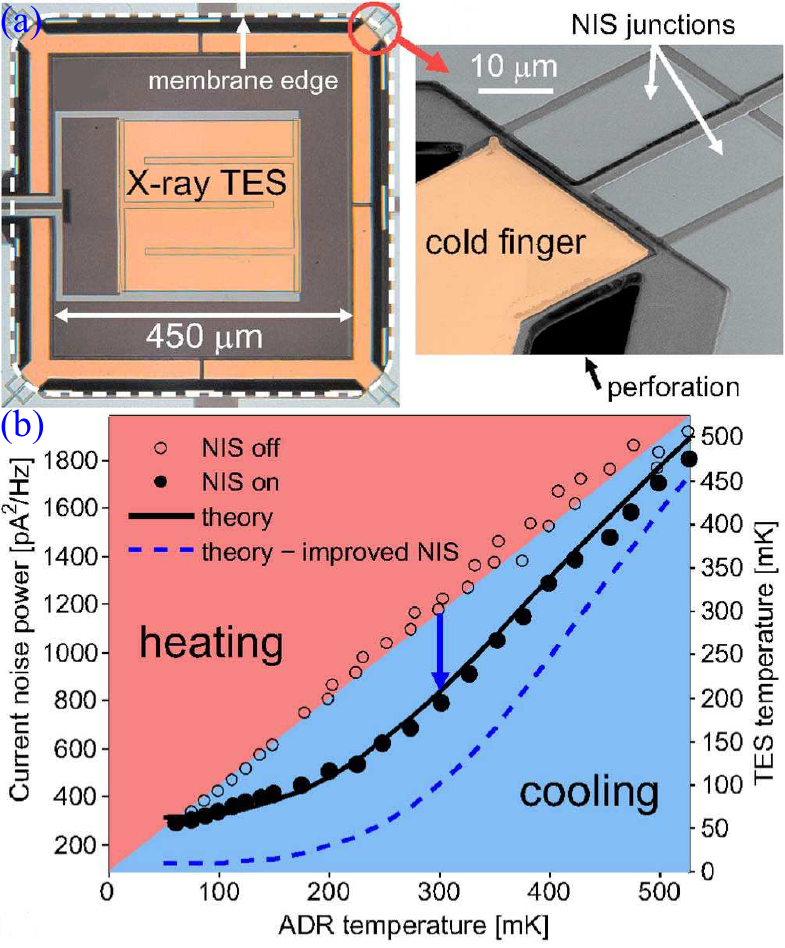}
\caption{Membrane cooler. 
\textbf{(a)} X-ray detector cooled with SINIS refrigerator. The detector is located on a silicon nitride membrane which has been perforated along the edges and is connected to the bulk substrate only through narrow bridges. The cooling junctions are located in the corners and the Y-shaped cold fingers extend to the membrane from the coolers.
\textbf{(b)} The noise properties of the X-ray detector, presented in (a), demonstrating the cooling effect.
\textit{Reprinted with permission from \cite{Miller2008}. Copyright 2008, American Institute of Physics.}
}
\label{fig:memcooler}
\end{figure}

The membrane cooling was first demonstrated by Luukanen et al. \cite{Luukanen2000} with a small membrane volume coupled to the bath through four few hundred micrometer long and $\sim$5 $\mu$m wide bridges. A considerable temperature decrease was achieved (from 200 mK to 100 mK), although the actual cooling power was modest ($\sim$ pW). However, actual application demonstrations have been done recently by the group of Ullom at NIST. They demonstrated first the cooling of a macroscopic size Ge cube \cite{Clark2005} and then an aluminium transition-edge detector, designed for X-ray detection \cite{Miller2008}. In the latter experiment, an effective temperature reduction from 300 mK to 190 mK was achieved in the noise properties of the detector, presenting a significant technological advance (see \fref{fig:memcooler}). The authors tested that inducing a 22 pW heating power to the membrane reduced the cooling by 7 mK, which would suggest an effective total cooling power of the order of few hundreds of pW.

Another geometry where the needed isolation of the phonon system from the environment can be achieved is in the form of a nanosized beam. The integration of other samples in the beam geometry is generally not very convenient but in these cases the beam itself can be the sample. In recent years, there has been much interest in the cooling down of local mechanical modes with the ultimate goal being the demonstration of the quantization of the mechanical vibrations of these mesoscopic objects \cite{OConnell2010,Teufel2011,Chan2011}. The problem here is that in order to demonstrate the quantization, these modes need to be very weakly coupled to the phonon bath of the bulk substrate (i.e. the $Q$-value of the resonator needs to be high). This condition makes the cooling mediated by the bulk phonon bath more difficult as there is inevitably some dissipated power generated by the measurement of the vibrations. Making the beam out of normal metal connected to NIS junctions would circumvent this problem as the local modes would then be directly cooled (through the electrons).

\begin{figure}
\centering
\includegraphics[width=1\textwidth]{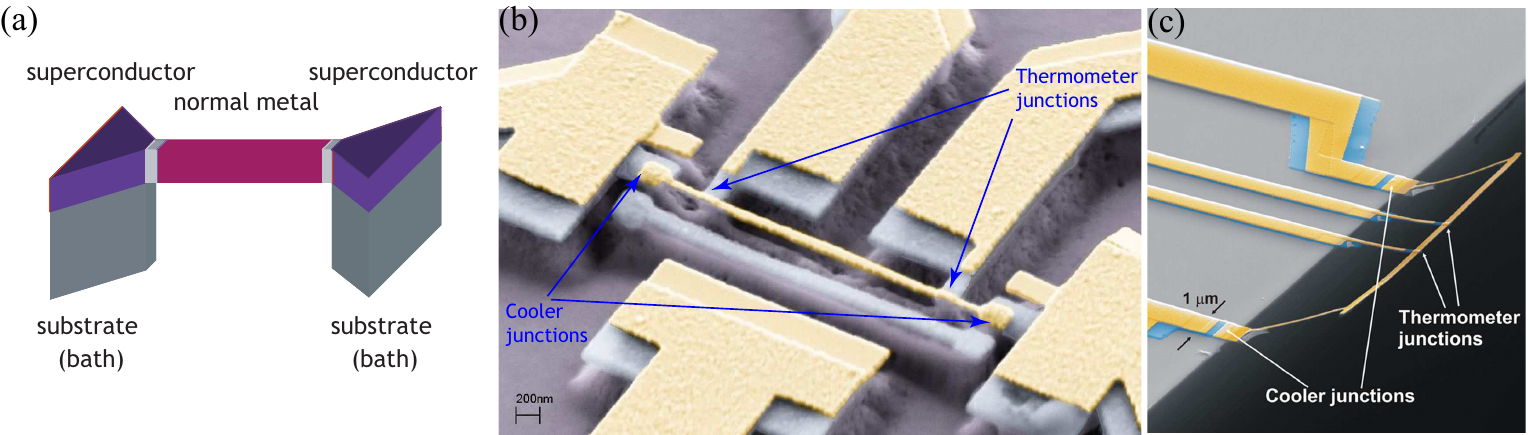}
\caption{Nanobeam coolers. 
\textbf{(a)} Schematic of the theoretical proposal made in \cite{Hekking2008}. A nanomechanical resonator made out of metal is placed in between two superconducting leads and tunnel coupled to them. These NIS contacts will cool the electrons of the beam and hence the mechanical vibrations through the $e-p$ coupling.
\textbf{(b)} A coloured SEM micrograph of the experimental realisation in \cite{Muhonen2009}. Successful electron cooling of the beam was demonstrated but the phonon temperature was not probed.
\textbf{(c)} Experimental realisation of \cite{Koppinen2009}. Here the thermal bottleneck are the long bridges extending to the beam and hence both the phonon and electron system of the beam are cooled.
\textit{(c) is a reprinted figure with permission from \cite{Koppinen2009}. Copyright 2009 by the American Physical Society.}
}
\label{fig:BeamCooler}
\end{figure}

This scenario was considered theoretically in \cite{Hekking2008} (see \fref{fig:BeamCooler} (a)). It had been experimentally verified already earlier that $e-p$ coupling can depend on the dimensionality of the phonon system \cite{Karvonen2007}. In \cite{Hekking2008}, authors showed that assuming a 1D phonon population and doing the conventional calculation \cite{Wellstood1994} for electron-phonon coupling yields a $T^3$ power law for the heat flow
\begin{equation}
P^{\rm 1D}_{e-ph} = \frac{\pi\zeta(3)}{6\zeta(5)}(\frac{\hbar c_l}{k_B})^2 \Sigma L (T_e^3 - T_{ph}^3),
\end{equation}
where $\zeta(3)/\zeta(5) \simeq 1.16$, $c_l$ is the speed of sound of the longitudinal modes, $\Sigma$ is the same electron-phonon coupling constant as in the 3D case and $L$ is the length of the beam. This coupling is between the electrons and the \textit{longitudinal} phonon modes, since in the first order perturbation theory the electrons do not couple to the transversal modes, i.e. the flexural modes of the beam. With this coupling, cooling the longitudinal modes of the resonator below the phonon bath temperature with NIS-cooler should be possible as long as the $Q$-factor of the resonator is above 100, an easy requirement to meet with mechanical resonators. The fabrication techniques needed for this kind of beam cooler were demonstrated in \cite{Muhonen2009} (see \fref{fig:BeamCooler} (b)), but the coupling between the electron system and the local mechanical modes remains an open question. Lately, many other proposals on cooling down the mechanical modes of a metallic resonator have also been reported \cite{Sonne2010, Sonne2011, Santandrea2011, Chamon2011}. These do not rely on NIS junctions.

In \cite{Koppinen2009}, a hybrid solution between the beam geometry and membrane geometry was fabricated (see \fref{fig:BeamCooler} (c)). Here the beam was made of silicon nitride and was connected to the bath only through narrow bridges. The thermal conductivity model is hence the same as in the earlier demonstrations of membrane cooling (including the cold fingers), but now the cooled phonon system is in the form of a beam. In this way, the authors were able to cool also the presumably 1D phonons of the beam but the experiment was not directly sensitive to any dimensionality or localization effects of the wire phonons. Nevertheless, the authors saw a power law of 2.8 at the lowest temperatures and attributed this to the 1D-2D phonon scattering at the bridge-bulk interface.
\section{SIS' coolers} \label{sec:SIS}

In all the NIS coolers presented in the previous section, the superconducting material used was aluminium (Al). As Al naturally forms a very high quality oxide layer when it is exposed to oxygen, this makes the fabrication of high quality tunnel barriers easy and has made Al by far the most popular superconducting material to use in the fabrication of tunnel barriers. However, when considering coolers, using some other materials with higher $T_C$ would have the obvious advantage of moving the optimum cooling temperature higher and providing higher cooling powers. This would be a very important step for technological applications. One might even envision a cascade cooler cooling from 4 K to below 100 mK utilizing different superconducting materials. So far, however, most attempts have been hindered by fabrication difficulties. The main problem lies in achieving low leakage junctions with other insulating barrier materials than aluminium oxide. One of the most common superconducting material is niobium (Nb), which has the advantage of having the highest $T_C$ ($\sim$ 9 K) among metallic elements and is also relatively easy to sputter deposit. However, achieving tunnel junctions with sufficiently low leakage currents has proved problematic with Nb and has so far prevented all demonstrations of significant cooling. Some progress has been recently made, however, by using aluminium nitride instead of aluminium oxide \cite{Zijlstra2007}.

A simple solution to this problem is to use the Al as a normal metal on top of which one can grow the oxide layer and then deposit a superconducting layer. This can be done and has been done by suppressing the superconductivity of Al with magnetic impurities, specifically manganese (Mn) \cite{Clark2004}. In \cite{Clark2004} the actual superconductor was, however, still Al. The approach had the advantage that the superconducting layer could be made arbitrarily thick (which is not possible when superconductor is deposited as the first layer due to fabrication technicalities). The thicker Al layer makes the quasiparticle overheating effects discussed in \sref{sec:diffModels} less harmful. This kind of Al-AlMn cooler was used in the membrane cooling demonstrations presented in \sref{sec:membr} \cite{Clark2005, Miller2008}. More recently, a trap layer was also introduced on top of the superconducting layer \cite{Oneill2011} in AlMn coolers. This means that the superconductor is then sandwiched between two normal metal films, and this seems to improve the cooling performance.

\begin{figure}
\centering
\includegraphics{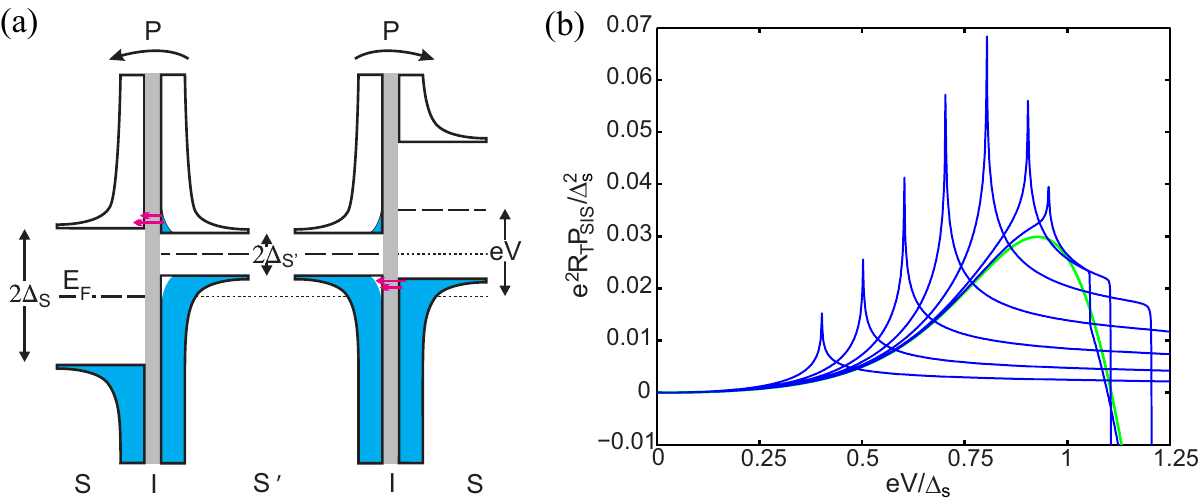}
\caption{SIS' cooler.
\textbf{(a)} The energy diagram of a SIS' structure. Basic principles are identical to \fref{fig:nis}. The qualitative differences are the different optimal cooling voltage and the divergent density of states when the two gap edges are aligned.
\textbf{(b)} Bias dependence of the cooling power of a SIS' junction. Smooth bottom line is the calculation for the corresponding SIN junction. The different lines are for different values of the energy gap of the second superconductor $\Delta_{S'}$. From left to right $\Delta_{S'}$ is 0.6, 0.5, 0.4, 0.3, 0.2, 0.1 and 0.05 times $\Delta_{S}$. The cooling power diverges at $eV = \Delta_S - \Delta_{S'}$.
}
\label{fig:SIS}
\end{figure}

Another solution is to use a different $T_c$ (and hence different energy gap) superconductor on top of an Al layer. This kind of SIS' (superconductor - insulator - superconductor with a different gap) cooler \cite{Frank1997} was first demonstrated with Ti/Al junctions \cite{Manninen1999}. The cooling power of a SIS' junction is in analogy to NIS cooling
\begin{eqnarray} \label{eq:PSIS}
P_\textrm{SIS'} =& \frac{1}{e^2R_T}\int \rmd E \, (E-eV) \, n_S(E) \, n_{S'}(E-eV) \nonumber \\
& \times [f_{S'}(E-eV)-f_S(E)].
\end{eqnarray}
\Eref{eq:PSIS} is very similar to \eref{eq:nisP} but has few important qualitative differences: (i) cooling power is maximised when the voltage is $(\Delta_S - \Delta_{S'})/e$, and (ii) at the optimum cooling point, the density of states diverges on both sides of the tunnel junction giving rise to nominally infinite cooling power. In \fref{fig:SIS}, we plot the normalised $P_\textrm{SIS'}$ as a function of $V$ for different values of $\Delta_{S'}$.

\begin{figure}
\centering
\includegraphics{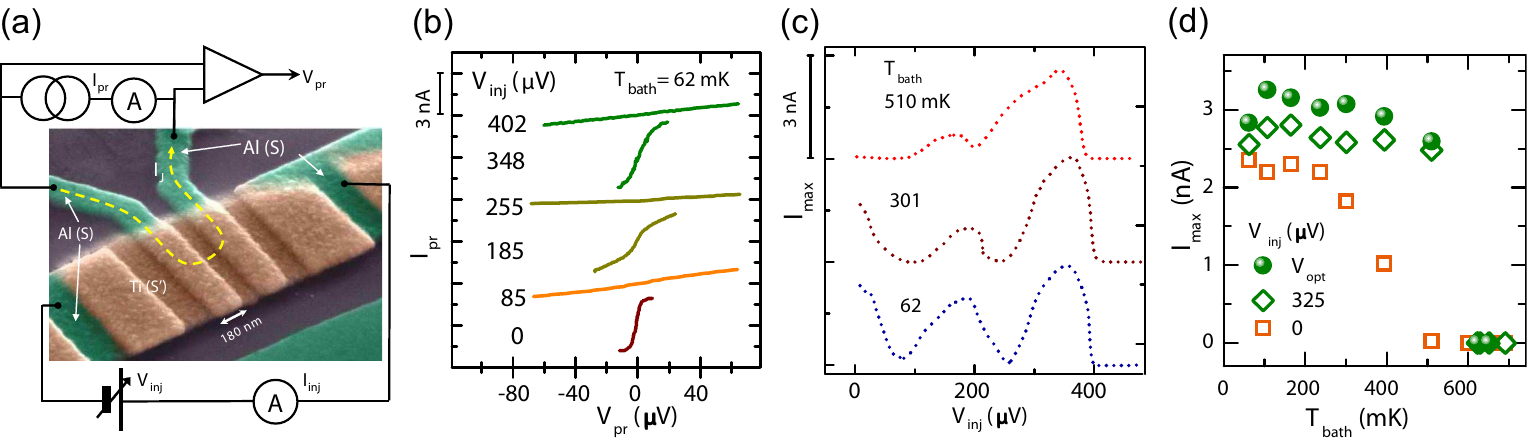}
\caption{SIS'IS cooling.
\textbf{(a)} The experimental structure. Ti island is connected to four Al leads through aluminium oxide tunnel barriers. Two of the junctions are used for cooling and the other two for thermometry.
\textbf{(b)} Current-voltage characteristics of the thermometer junctions measured with different biasing of the cooling junctions. The critical current of the junctions is seen to first decrease, then increase, decrease and increase again, before the supercurrent finally quenches at high biases.
\textbf{(c)} The thermometer critical current seen at different bath temperatures as a function of the biasing of the cooler junctions. Increase in the maximum supercurrent corresponds to cooling of the quasiparticle system of the Ti island. A double peak structure is seen (see text).
\textbf{(d)} The maximum thermometer supercurrent as a function of the bath temperature at zero cooler bias, optimal bias and one point in between.
\textit{Reprinted figure with permission from \cite{Tirelli2008}. Copyright 2008 by the American Physical Society.}
}
\label{fig:Tirelli}
\end{figure}

The SINIS analogy was later found to be too simplistic \cite{Tirelli2008} for the case of a SIS'IS structure. When applying a current over a superconducting tunnel junction, the voltage will not in general develop gradually but rather, there is an abrupt change from the supercurrent state to a quasiparticle current state when the current exceeds the critical current of the junction. This is commonly known as switching of a Josephson junction and will lead to voltage $\Delta/e$ to abruptly appear over the junction. As two tunnel junctions will never be exactly the same in a double tunnel junction structure, one junction will tend to switch with lower current and develop a voltage over it. When more current is then applied over the structure, the voltage across the first junction will increase until the other junction also switches. This takes the voltage $\sim \Delta/e$ over it, causing the voltage over the first junction to decrease. When the first switching happens, the voltage reaches the difference of the two gaps $V \approx (\Delta_S - \Delta_{S'})/e$. The other junction is then driven to the cooling regime and a drop in S' quasiparticle temperature can be seen (see \fref{fig:Tirelli} from \cite{Tirelli2008}). With increasing bias, the sharp cooling peak will turn into heating until the other junction also switches, at which point a second cooling peak is seen at $V \approx 2(\Delta_S - \Delta_{S'})/e$. 

In \cite{Tirelli2008} the authors used the critical current of two additional SIS' probe junctions attached to the cooled Ti island as thermometers. With the SIS'IS cooler biased at the optimum point, they could reach a critical current on the probe junctions at 0.5 K that exceeded the critical current they could see at the lowest cryostat temperature ($\sim 50$ mK) without biasing the cooler. This suggests that the S' island quasiparticle population was cooled from 0.5 K to below what it could be cooled through the phonon system.  

A fully Al SIS' cooler was presented in \cite{Ferguson2008}, where the difference in energy gaps was achieved using Al layers with different thickness. The energy gap of aluminium increases when the thickness of the metal layer is below few tens of nanometers. In \cite{Ferguson2008} Al layers of 30 and 10 nm were used, giving energy gaps of 209 and 250 $\mu$eV, respectively. The SIS' structure was then used to cool down one of the electrodes of a single Cooper pair transistor (SCPT). SCPT is essentially a single electron transistor working with Cooper pairs instead of single electrons. SCPTs are, however, prone to quasiparticle poisoning \cite{Tuominen1992} causing the devices to have 1e periodicity in their gate voltage characteristics in addition to the expected 2e. By biasing the SIS' structure which then cooled one of the leads of a SCPT, the author was able to show that the probability of having unpaired electrons on the island went down by a factor of two.

\begin{figure}
\centering
\includegraphics{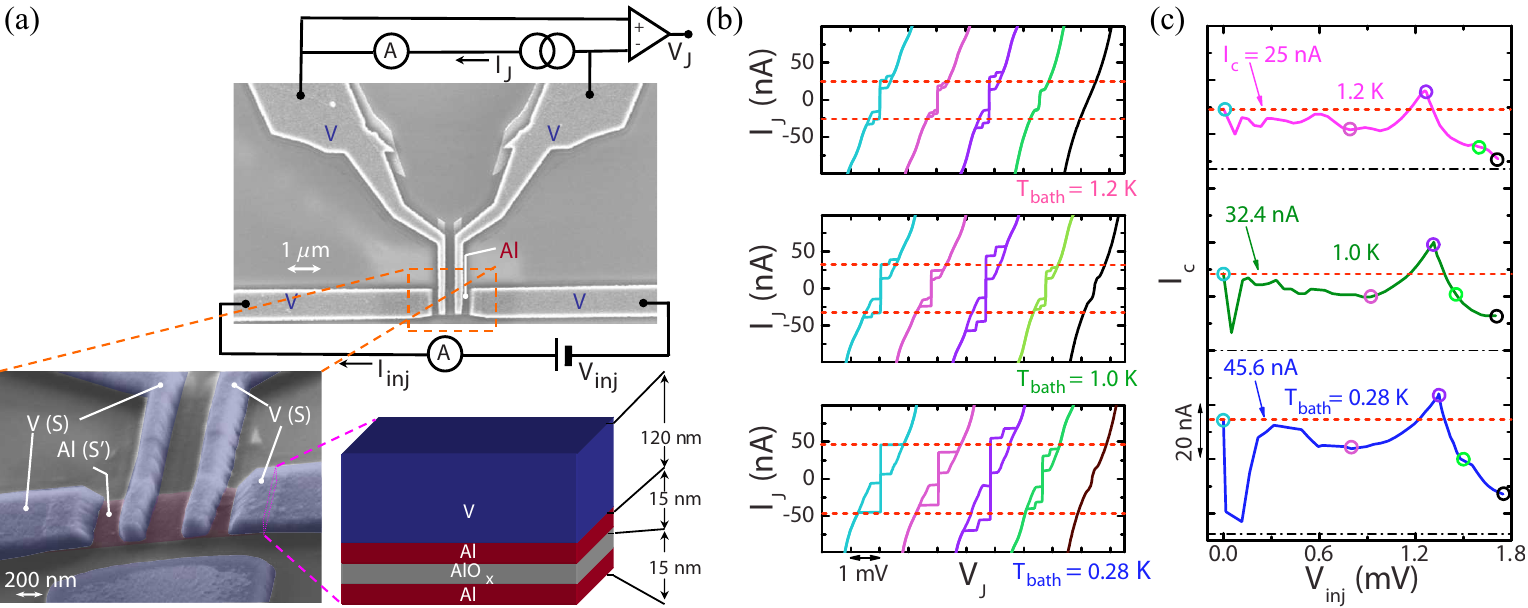}
\caption{Vanadium cooler.
\textbf{(a)} The cooler structure. An Al island is connected to four V leads, two for thermometry and two for cooling. Above the aluminium oxide a thin layer of Al is deposited before the V deposition in order to protect the oxide layer.
\textbf{(b)} Thermometer current-voltage characteristics at different values of the cooler biasing. From left to right the curves refer to biasing points shown by circles in (c).
\textbf{(c)} The supercurrent of the thermometer junctions as a function of the cooler biasing. A clear cooling peak can be seen at $V \sim 1.2$ mV. In the middle curve ($T_{\rm bath} = 1$ K), the increase in critical current corresponds to quasiparticle temperature of 0.4 K at the Al island.
\textit{Reprinted with permission from \cite{Quaranta2011}. Copyright 2011, American Institute of Physics.}
}
\label{fig:quaranta}
\end{figure}

None of the aforementioned coolers addressed the cooling from higher temperatures that was given as a motivation at the beginning of the section. Lately significant progress to this direction has been made in \cite{Quaranta2011} where the authors succeeded in making a vanadium (V, $T_c \sim 5$ K) - aluminium SIS' cooler (see \fref{fig:quaranta}). In order to make a good quality junction it was necessary to cover first the oxide layer with a small amount of Al before depositing V. The Al/V bilayer had a $T_c$ of 4 K and the authors were able to achieve a significant temperature reduction as deduced from the critical current of two additional probe junctions: they cooled the quasiparticle system of Al from 1 K to 400 mK.

\section{Schottky barrier coolers} \label{sec:sm}

\begin{figure}
\centering
\includegraphics{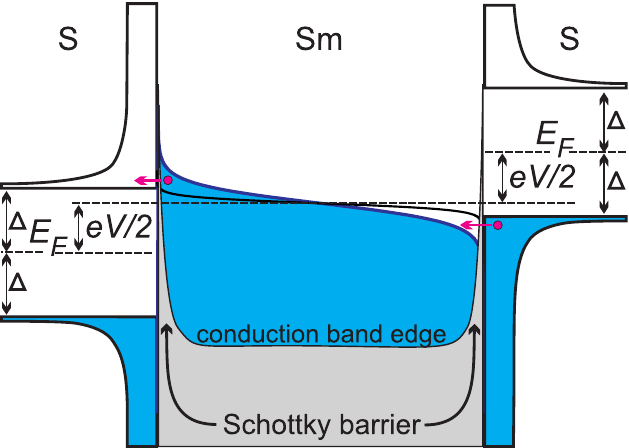}
\caption{The energy diagram of an S-Sm-S structure. Basic principles are identical to those in \fref{fig:nis}. The insulating layers are replaced by a Schottky barrier at the Al-Si interface. Energy gap of the semiconductor plays no role, as the island is degenerately doped.
}
\label{fig:Schottky}
\end{figure}

The basic principles of NIS cooling apply also if the normal metal island is replaced by a heavily doped semiconductor (see \fref{fig:Schottky} (a)). The superconductor - semiconductor (S-Sm) cooler presents some benefits compared to normal metals:
(i) The electron-phonon coupling strength is generally weaker in semiconductors than in normal metals (at 100 mK, Si, depending on the doping level, has roughly 1-2 orders of magnitude smaller $e-p$ coupling than Cu) and
(ii) the Schottky barrier can play the role of the tunnelling barrier and hence no oxide layer is needed between the superconductor and the semiconductor. This makes fabrication of especially large area junctions more straightforward than with the standard shadow evaporation techniques.
In addition, both the Schottky barrier resistance and the electron-phonon coupling can be tuned by varying the doping level of the semiconducting island. The most obvious drawbacks are that even highly doped semiconductors have a higher resistivity than metals and hence more parasitic Joule heating is generated. Furthermore, relatively large subgap currents are typically observed leading to non-ideal cooler performance.

The cooling effect in S-Sm structures was first presented in \cite{Savin2001} and extended in \cite{Savin2003, Savin2004}. In \cite{Savin2001}, a cooling power of roughly 0.5 pW was achieved with two 5x18 $\mu$m$^2$ junctions having total $R_T$ of 800 $\Omega$. This lead to 30 \% drop in temperature from 175 mK because of the small $e-p$ coupling. The doping level of the n$^+$ silicon was $4\times10^{19}$ cm$^{-3}$. In \cite{Savin2003, Savin2004}, the work was extended to multiple n$^+$ doping levels of the semiconducting island. It was found that, in agreement with the theory, the contact resistance $R_T$ of Al-Si interface scaled as $\exp(N^{-1/2})$ where $N$ is the doping level. For the cooling effect, this is partly compensated by the increase in the $e-p$ coupling due to higher doping. However, the latter effect was found to be only linearly proportional to doping and hence larger doping should lead to increase in cooling power. Yet the larger cooling effect was seen only at higher temperatures (above $\sim$300 mK) and increasing doping to above $1\times10^{20}$ cm$^{-3}$ made the cooling effect smaller. This was attributed to large ohmic leakage currents through the barrier at lower transparencies, i.e. effectively the $\gamma$ parameter in \eref{eq:DOSdynes}. The $\gamma$ generally found in Al-Si junctions has been $10^{-2}-10^{-1}$, which is a few orders of magnitude worse than in Al-Al$_2$O$_3$-Cu junctions. In \cite{Buonomo2003}, also niobium-silicon junctions were studied. The basic IVs of SINIS structures could be reproduced also with Nb, but no cooling was observed. This was again due to large subgap leakage currents. The contact resistance between Nb and Si was found to be much smaller than with Al and Si, in accordance to expectations as the Schottky barrier height is smaller in this case.

In \cite{Savin2001, Savin2003, Savin2004}, $e-p$ coupling in Si was modelled with the $T^5$ power law as in normal metal case. However, later it was confirmed to follow a higher $T^6$ law \cite{Kivinen2003}. Theoretically, this power law was expected for (single-valley) semiconductors in two dimensions at the diffusive limit \cite{Sergeev2005} but the theoretical prefactor was several orders of magnitude smaller than measured. In \cite{Prunnila2005, Prunnila2007}, the fact that multiple conduction band valleys exist in Si was included into the theoretical analysis. Phonons can lift the degeneracy between the different valleys so that the valley degree of freedom starts to play a role in the low temperature $e-p$ coupling. This was shown to lead to the $T^6$ power law and, because this channel is unscreened at low temperatures, have a prefactor of the correct order of magnitude with experiments. The prefactor and temperature dependence were also experimentally confirmed in \cite{Prunnila2005}.

\begin{figure}
\centering
\includegraphics[width=0.8\textwidth]{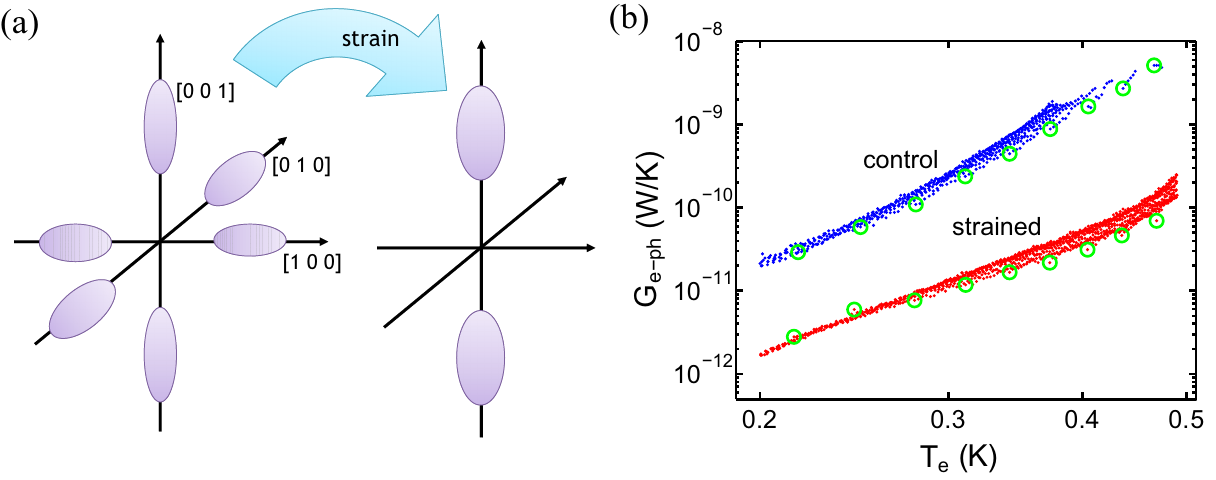}
\caption{Effects of strain induced by lattice mismatch to the $e-ph$ coupling in n$^+$Si.
\textbf{(a)} Energy diagram of the conduction band valleys and how the strain should affect them. The perpendicular valleys are lifted in energy and hence become depopulated.
\textbf{(b)} Experimental results from \cite{Muhonen2011}. The $G_{e-ph} = \partial P_{e-ph}/\partial T$ is smaller in the strained sample, as compared to an unstrained control sample, by a factor of about 20.
}
\label{fig:strain}
\end{figure}

Recently \cite{Muhonen2011, Prest2011}, it has been tested how removing this relaxation channel will modify the $e-p$ relaxation in Si. Similarly as phonons lift the degeneracy between the different valleys, one can also lift the degeneracy ``permanently'' by inducing strain to the silicon layer. If the strain induced energy splitting is larger than the Fermi energy (as measured from the bottom of the conduction band), the in-plane valleys will depopulate and hence the screened, single-valley case prevails. Theoretically, the $e-p$ coupling should decrease by several orders of magnitude \cite{Prunnila2007}. The effect was tested experimentally in \cite{Muhonen2011} (see \fref{fig:strain}). The authors found that the $e-p$ coupling was indeed lower in the strained sample as compared to an unstrained control sample and to previous experiments, although the reduction factor was only between one and two orders of magnitude. Nevertheless, the heat flow from the phonon system decreases significantly and it can be useful for cooler applications, as was demonstrated in \cite{Prest2011} where enhanced electron cooling in strained silicon sample was seen.

\section{Quantum dot refrigerator}

\begin{figure}
\centering
\includegraphics{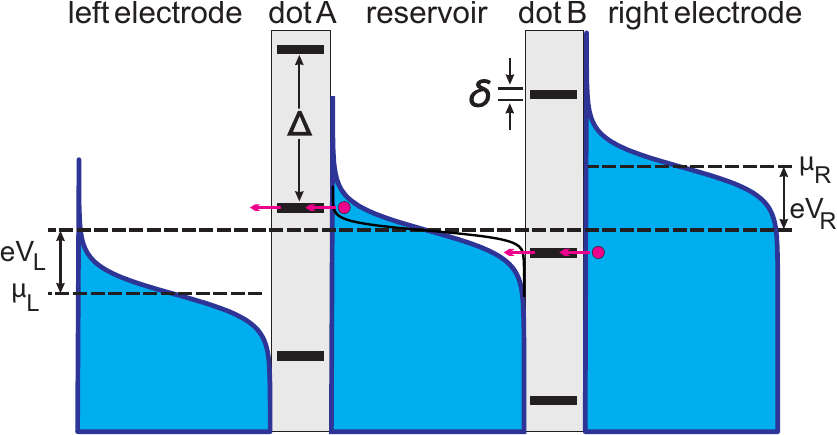}
\caption{Energy diagram of a quantum dot cooler. The two quantum dots act as filters allowing only electrons of certain energy to tunnel from the central reservoir. Proper voltage bias applied over the structure cools the electron system of the central reservoir.
}
\label{fig:qdr}
\end{figure}

Refrigeration using a semiconducting quantum dot, instead of a metal hybrid, had been proposed by Edwards \textit{et al.} \cite{Edwards1993,Edwards1995}. The principle of such a quantum dot refrigerator (QDR) is shown in \fref{fig:qdr}. The proposed device consisted of a central island which is separated from the leads by two quantum dots, A and B. The dots have an energy level separation of $\Delta$ and the dot energies, $E_A$ and $E_B$, can be adjusted by separate gate voltages. It is assumed that the dots are tunnel coupled both to the leads and to the island but no coupling exists directly between the leads and the island. If one applies a small dc voltage across the leads so that the energy separation between the chemical potentials $\mu_L$ and $\mu_R$ is smaller than $2\Delta$, the chemical potential of the island $\mu_0$ will lie midway between $\mu_L$ and $\mu_R$. There will be then exactly one or zero energy levels between $\mu_0$ and the chemical potentials of either lead. By positioning the energy level $E_A$ ($E_B$) so that it is slightly above (below) $\mu_0$, an energy $E_B-E_A$ is removed from the central area as an electron travels from one lead to another. For this process to be energetically possible, the separations between the energy levels of the quantum dots and the island must be of the order of $k_BT$.

\begin{figure}
\centering
\includegraphics{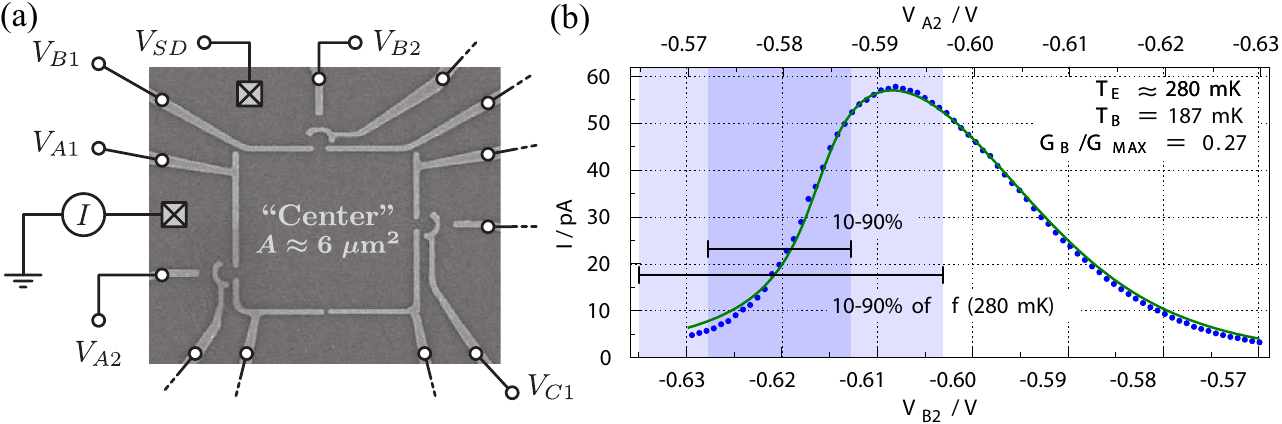}
\caption{Quantum dot cooler.
\textbf{(a)} The experimental structure. Electrodes on top of the 2DEG define the quantum dots and the central island. $V_{\rm A2}$ and $V_{\rm B2}$ are the gate voltages controlling the energy levels of the quantum dots.
\textbf{(b)} The current through the device as a function of $V_{\rm A2}$ and $V_{\rm B2}$, demonstrating the cooling effect. With equilibrium electron distribution the current curve would be symmetric. The change in width on the left side of the curve is a result of cooling. Authors inferred from this asymmetry a temperature reduction of 93 mK starting from 280 mK.
\textit{Reprinted figure with permission from \cite{Prance2009}. Copyright 2009 by the American Physical Society.}
}
\label{fig:prance}
\end{figure}

The concept was experimentally tested and demonstrated recently by Prance \textit{et al.} \cite{Prance2009} in a 2D electron gas (2DEG), which is also the system Edwards \textit{et al.} originally proposed. 2DEGs have a very weak coupling to the acoustic phonons of the lattice which both makes cooling them by conventional methods hard and facilitates significant changes to the electron temperature with even modest cooling powers by direct electronic cooling. In the experiment of \cite{Prance2009}, a 6 $\mu$m$^2$ central area of 2DEG was cooled from 280 mK down to 187 mK under optimized bias conditions of the device, see \fref{fig:prance}. The data are consistent with a thermal model (quasi-equilibrium) down to about 120 mK bath temperature. Below that the cooling becomes ineffective and the data cannot be fit to a simple model where conductance is parametrized by temperature. The authors ascribe this to poor electron-electron relaxation at low temperatures leading to a non-equilibrium energy distribution. 

More recently \cite{Gasparinetti2011}, local thermometry of the 2DEG reservoir was demonstrated with a similar structure. The temperature of the reservoir was deduced from the thermally broadened conductance of a quantum dot, allowing the measurement of electron-phonon coupling constant of the 2DEG. An alternative cooler mechanism for 2DEG has been proposed in \cite{Giazotto2007}.

\section{Perspectives}

It has been 45 years since sub-100 mK temperatures were first opened up to researchers by the advent of dilution refrigerators. Although this technique has steadily matured and proven to be a very important workhorse for low temperature research, it remains complicated and is getting increasingly expensive due to the unstable price of $^3$He. A significant demand exists for alternative techniques that would not require users to be experts in low temperature techniques and, ideally, would not use cryoliquids. The micron-scale coolers presented in this review have the potential to provide these advantages in the future, although a significant amount of research is still needed. This goal could be accomplished, for example, by combining already available commercial pulsed cryocoolers (which can cool to $\sim 4$ K) with a micron-scale solid state cooler based on e.g. a cascade of superconducting coolers with different superconducting materials.

\begin{figure}
\centering
\includegraphics[height = 6cm]{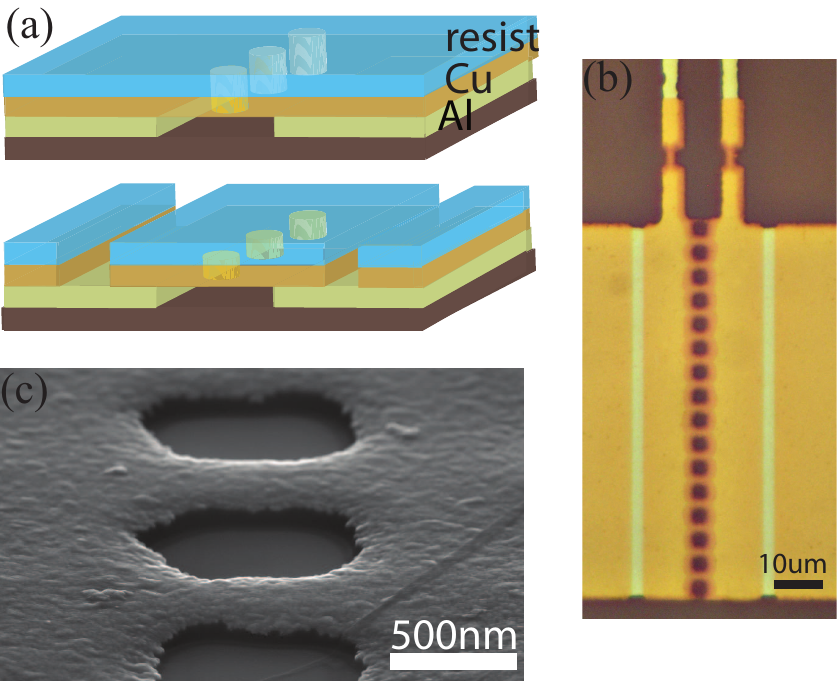}
\caption{Fabrication process for large area junctions presented in \cite{Nguyen2011}.
\textbf{(a)} Process flow: Al-Al$_2$O$_3$-Cu structure is evaporated on a substrate and covered with resist. Dots are patterned to the resist (with photolithography) and etched through the copper with ion-beam-etcher. Aluminium is wet etched through the holes, creating a suspended Cu structure in the middle. Finally, in a second lithography (EBL) step, the cooler junctions are defined by cutting through the Cu at a distance of few micrometers from the holes. With this technique very large area junctions can be achieved and the thickness of the Al layer is not limited (facilitating easier removal of hot quasiparticles).
\textbf{(b)} Optical microscope image of a cooler structure. Two junctions at the top of the picture are for thermometry. The large cooling junctions are defined by the cut in copper (showing as light vertical stripes) and the cut in Al in the middle.
\textbf{(c)} View of the middle normal metal part, showing that it is freely suspended.
}
\label{fig:hung}
\end{figure}

Important aspects of the development process of the micron-scale coolers, which have been mostly not mentioned in this review, are the fabrication techniques. The overwhelming majority of superconducting coolers demonstrated so far have been fabricated with e-beam lithography (EBL) combined with multiple angle evaporation. Although a very useful technique in laboratory settings, this can hardly be considered an industrial process and does place severe limitations of the junction sizes (and hence the cooling powers) that can be produced. Larger Al-Al$_2$O$_3$-Cu junctions were demonstrated already ten years ago with mechanical masks combined with multiple angle evaporation \cite{Pekola2000a}. Another possibility raised early on is to use degenerately doped semiconductor instead of normal metal (see \sref{sec:sm}). However, more recently, demonstrations have been made with photolithography. A prime example are the AlMn based coolers presented in \sref{sec:SIS} and used in the membrane cooling demonstrations (see \sref{sec:membr}). Another very recent example is presented in \fref{fig:hung}. NIS coolers can now achieve effective cooling power of hundreds of picowatts, which is a considerable increase from the sub-picowatt range of the original demonstrations, but there is still a lot of room for improvement.

It has, however, become increasingly clear that the generalisation of superconducting coolers to high cooling powers is not as straightforward as one might think. The high density of non-equilibrium quasiparticles created means that thermalisation of the ``backside'' of the cooler and local phonon heating become significant concerns which degrade the efficiency of the cooler at high cooling powers. These issues must be addressed in designing any high power cooler. The solutions presented so far rely on quasiparticle traps (see \sref{sec:diffModels}) and separating the cooled phonon system from the phonons in contact with the cooler (perforated membranes as in \sref{sec:membr}). As was mentioned in \sref{sec:magn}, it was also recently found out that small magnetic fields can help in this respect.

On the other hand, in some applications, the high cooling power is not important but rather the goal is to reach as low as possible electronic temperatures. In this regime, significant progress has also been made lately, as the significance of the coupling of an electronic conductor to its electromagnetic environment has become clear in this context (see \sref{sec:photons}). This has also helped to increase the understanding of the ultimate limitations of the quality of NIS junctions.

Recently, several new kinds of micron-scale coolers have been proposed. Partly this has been driven by the explosion of interest into cooling of the local mechanical modes of nanomechanical resonators. This is good news for the field and shows that for the micron-scaled coolers, there remains a lot of research to do, both for finding totally new directions as well as in improving the known ones.

\ack
We wish to thank T. Aref for proofreading the manuscript and V. F. Maisi, J.T. Peltonen, M. Prunnila and A.V. Timofeev for providing material for the article. We acknowledge financial support by EPSRC grant EP/F040784/1 and the European Community's FP7 Programme under Grant Agreements No. 228464 (MICROKELVIN, Capacities Specific Programme).

\section*{References}
\bibliographystyle{unsrt}
\bibliography{RPP_review_refs}

\begin{thebibliography}{10}

\bibitem{Giazotto2006}
Francesco Giazotto, Tero~T. Heikkil\"a, Arttu Luukanen, Alexander~M. Savin, and
  Jukka~P. Pekola.
\newblock Opportunities for mesoscopics in thermometry and refrigeration:
  Physics and applications.
\newblock {\em Rev. Mod. Phys.}, 78(1):217--274, Mar 2006.

\bibitem{Prance2009}
J.~R. Prance, C.~G. Smith, J.~P. Griffiths, S.~J. Chorley, D.~Anderson,
  G.~A.~C. Jones, I.~Farrer, and D.~A. Ritchie.
\newblock Electronic refrigeration of a two-dimensional electron gas.
\newblock {\em Phys. Rev. Lett.}, 102(14):146602, Apr 2009.

\bibitem{Pekola2004}
J.~P. Pekola, T.~T. Heikkil\"a, A.~M. Savin, J.~T. Flyktman, F.~Giazotto, and
  F.~W.~J. Hekking.
\newblock Limitations in cooling electrons using normal-metal-superconductor
  tunnel junctions.
\newblock {\em Phys. Rev. Lett.}, 92(5):056804, Feb 2004.

\bibitem{Pothier1997}
H.~Pothier, S.~Gu\'eron, Norman~O. Birge, D.~Esteve, and M.~H. Devoret.
\newblock Energy distribution function of quasiparticles in mesoscopic wires.
\newblock {\em Phys. Rev. Lett.}, 79(18):3490--3493, Nov 1997.

\bibitem{Rajauria2007}
Sukumar Rajauria, P.~S. Luo, T.~Fournier, F.~W.~J. Hekking, H.~Courtois, and
  B.~Pannetier.
\newblock Electron and phonon cooling in a
  superconductor--normal-metal--superconductor tunnel junction.
\newblock {\em Phys. Rev. Lett.}, 99(4):047004, Jul 2007.

\bibitem{Wellstood1994}
F.~C. Wellstood, C.~Urbina, and John Clarke.
\newblock Hot-electron effects in metals.
\newblock {\em Phys. Rev. B}, 49(9):5942--5955, Mar 1994.

\bibitem{Visser2011}
P.~J. de~Visser, J.~J.~A. Baselmans, P.~Diener, S.~J.~C. Yates, A.~Endo, and
  T.~M. Klapwijk.
\newblock Number fluctuations of sparse quasiparticles in a superconductor.
\newblock {\em Phys. Rev. Lett.}, 106(16):167004, Apr 2011.

\bibitem{Palmer2007}
B.~S. Palmer, C.~A. Sanchez, A.~Naik, M.~A. Manheimer, J.~F. Schneiderman,
  P.~M. Echternach, and F.~C. Wellstood.
\newblock Steady-state thermodynamics of nonequilibrium quasiparticles in a
  cooper-pair box.
\newblock {\em Phys. Rev. B}, 76(5):054501, Aug 2007.

\bibitem{Shaw2008}
M.~D. Shaw, R.~M. Lutchyn, P.~Delsing, and P.~M. Echternach.
\newblock Kinetics of nonequilibrium quasiparticle tunneling in superconducting
  charge qubits.
\newblock {\em Phys. Rev. B}, 78(2):024503, Jul 2008.

\bibitem{Martinis2009}
John~M. Martinis, M.~Ansmann, and J.~Aumentado.
\newblock Energy decay in superconducting josephson-junction qubits from
  nonequilibrium quasiparticle excitations.
\newblock {\em Phys. Rev. Lett.}, 103(9):097002, Aug 2009.

\bibitem{Barends2008}
R.~Barends, J.~J.~A. Baselmans, S.~J.~C. Yates, J.~R. Gao, J.~N. Hovenier, and
  T.~M. Klapwijk.
\newblock Quasiparticle relaxation in optically excited high-$q$
  superconducting resonators.
\newblock {\em Phys. Rev. Lett.}, 100(25):257002, Jun 2008.

\bibitem{Saira2011}
O.~P. {Saira}, A.~{Kemppinen}, V.~F. {Maisi}, and J.~P. {Pekola}.
\newblock Is aluminum a perfect superconductor?
\newblock {\em ArXiv e-prints}, 1106.1326v2, Jun 2011.

\bibitem{Arutyunov2011}
K.~Yu. Arutyunov, H.-P. Auraneva, and A.~S. Vasenko.
\newblock Spatially resolved measurement of nonequilibrium quasiparticle
  relaxation in superconducting al.
\newblock {\em Phys. Rev. B}, 83(10):104509, Mar 2011.

\bibitem{Kaplan1976}
S.~B. Kaplan, C.~C. Chi, D.~N. Langenberg, J.~J. Chang, S.~Jafarey, and D.~J.
  Scalapino.
\newblock Quasiparticle and phonon lifetimes in superconductors.
\newblock {\em Phys. Rev. B}, 14(11):4854--4873, Dec 1976.

\bibitem{Timofeev2009}
A.~V. Timofeev, C.~Pascual Garc\'\i{}a, N.~B. Kopnin, A.~M. Savin, M.~Meschke,
  F.~Giazotto, and J.~P. Pekola.
\newblock Recombination-limited energy relaxation in a
  bardeen-cooper-schrieffer superconductor.
\newblock {\em Phys. Rev. Lett.}, 102(1):017003, Jan 2009.

\bibitem{Bardeen1959}
J.~Bardeen, G.~Rickayzen, and L.~Tewordt.
\newblock Theory of the thermal conductivity of superconductors.
\newblock {\em Phys. Rev.}, 113(4):982--994, Feb 1959.

\bibitem{Virtanen2007}
P.~Virtanen and T.T. Heikkil\"a.
\newblock Thermoelectric effects in superconducting proximity structures.
\newblock {\em Applied Physics A: Materials Science \& Processing},
  89:625--637, 2007.

\bibitem{Peltonen2010}
J.~T. Peltonen, P.~Virtanen, M.~Meschke, J.~V. Koski, T.~T. Heikkil\"a, and
  J.~P. Pekola.
\newblock Thermal conductance by the inverse proximity effect in a
  superconductor.
\newblock {\em Phys. Rev. Lett.}, 105(9):097004, Aug 2010.

\bibitem{Ojanen2007}
Teemu Ojanen and Tero~T. Heikkil\"a.
\newblock Photon heat transport in low-dimensional nanostructures.
\newblock {\em Phys. Rev. B}, 76:073414, Aug 2007.

\bibitem{Ojanen2008}
Teemu Ojanen and Antti-Pekka Jauho.
\newblock Mesoscopic photon heat transistor.
\newblock {\em Phys. Rev. Lett.}, 100:155902, Apr 2008.

\bibitem{Pascal2011}
L.~M.~A. Pascal, H.~Courtois, and F.~W.~J. Hekking.
\newblock Circuit approach to photonic heat transport.
\newblock {\em Phys. Rev. B}, 83:125113, Mar 2011.

\bibitem{Pendry1983}
J~B Pendry.
\newblock Quantum limits to the flow of information and entropy.
\newblock {\em Journal of Physics A: Mathematical and General}, 16(10):2161,
  1983.

\bibitem{Schmidt2004}
D.~R. Schmidt, R.~J. Schoelkopf, and A.~N. Cleland.
\newblock Photon-mediated thermal relaxation of electrons in nanostructures.
\newblock {\em Phys. Rev. Lett.}, 93(4):045901, Jul 2004.

\bibitem{Meschke2006}
Matthias Meschke, Wiebke Guichard, and Jukka~P. Pekola.
\newblock Single-mode heat conduction by photons.
\newblock {\em Nature}, 444(7116):187--190, Nov 2006.

\bibitem{Timofeev2009a}
Andrey~V. Timofeev, Meri Helle, Matthias Meschke, Mikko M\"ott\"onen, and
  Jukka~P. Pekola.
\newblock Electronic refrigeration at the quantum limit.
\newblock {\em Phys. Rev. Lett.}, 102(20):200801, May 2009.

\bibitem{Pekola2010}
J.~P. Pekola, V.~F. Maisi, S.~Kafanov, N.~Chekurov, A.~Kemppinen, Yu.~A.
  Pashkin, O.-P. Saira, M.~M\"ott\"onen, and J.~S. Tsai.
\newblock Environment-assisted tunneling as an origin of the dynes density of
  states.
\newblock {\em Phys. Rev. Lett.}, 105(2):026803, Jul 2010.

\bibitem{Barends2011}
R.~Barends, J.~Wenner, M.~Lenander, Y.~Chen, R.~C. Bialczak, J.~Kelly,
  E.~Lucero, P.~O'Malley, M.~Mariantoni, D.~Sank, H.~Wang, T.~C. White, Y.~Yin,
  J.~Zhao, A.~N. Cleland, John~M. Martinis, and J.~J.~A. Baselmans.
\newblock Minimizing quasiparticle generation from stray infrared light in
  superconducting quantum circuits.
\newblock {\em Applied Physics Letters}, 99(11):113507, 2011.

\bibitem{Paik2011}
H.~{Paik}, D.~I. {Schuster}, L.~S. {Bishop}, G.~{Kirchmair}, G.~{Catelani},
  A.~P. {Sears}, B.~R. {Johnson}, M.~J. {Reagor}, L.~{Frunzio}, L.~{Glazman},
  S.~M. {Girvin}, M.~H. {Devoret}, and R.~J. {Schoelkopf}.
\newblock {Observation of high coherence in Josephson junction qubits measured
  in a three-dimensional circuit QED architecture}.
\newblock {\em ArXiv e-prints, accepted to Phys. Rev. Lett}, 1105.4652v4, May
  2011.

\bibitem{Anghel2001}
D.~V. Anghel and J.~P. Pekola.
\newblock Noise in refrigerating tunnel junctions and in microbolometers.
\newblock {\em Journal of Low Temperature Physics}, 123:197--218, 2001.

\bibitem{Mueller1997}
Heinz-Olaf M\"uller and K.~A. Chao.
\newblock Electron refrigeration in the tunneling approach.
\newblock {\em Journal of Applied Physics}, 82(1):453--456, 1997.

\bibitem{Dubos2001}
P.~Dubos, H.~Courtois, B.~Pannetier, F.~K. Wilhelm, A.~D. Zaikin, and
  G.~Sch\"on.
\newblock Josephson critical current in a long mesoscopic s-n-s junction.
\newblock {\em Phys. Rev. B}, 63:064502, Jan 2001.

\bibitem{Heikkila2002}
Tero~T. Heikkil\"a, Jani S\"arkk\"a, and Frank~K. Wilhelm.
\newblock Supercurrent-carrying density of states in diffusive mesoscopic
  josephson weak links.
\newblock {\em Phys. Rev. B}, 66:184513, Nov 2002.

\bibitem{Jiang2003}
Z.~Jiang, H.~Lim, V.~Chandrasekhar, and J.~Eom.
\newblock Local thermometry technique based on proximity-coupled
  superconductor/normal-metal/superconductor devices.
\newblock {\em Applied Physics Letters}, 83(11):2190--2192, 2003.

\bibitem{Courtois2008}
H.~Courtois, M.~Meschke, J.~T. Peltonen, and J.~P. Pekola.
\newblock Origin of hysteresis in a proximity josephson junction.
\newblock {\em Phys. Rev. Lett.}, 101:067002, Aug 2008.

\bibitem{Meschke2009}
M.~Meschke, J.~Peltonen, H.~Courtois, and J.~Pekola.
\newblock Calorimetric readout of a superconducting proximity-effect
  thermometer.
\newblock {\em Journal of Low Temperature Physics}, 154:190--198, 2009.

\bibitem{Ullom1998}
J.~N. Ullom, P.~A. Fisher, and M.~Nahum.
\newblock Energy-dependent quasiparticle group velocity in a superconductor.
\newblock {\em Phys. Rev. B}, 58(13):8225--, Oct 1998.

\bibitem{Rajauria2009}
Sukumar Rajauria, Herv\&eacute; Courtois, and Bernard Pannetier.
\newblock Quasiparticle-diffusion-based heating in superconductor tunneling
  microcoolers.
\newblock {\em Phys. Rev. B}, 80(21):214521--, Dec 2009.

\bibitem{Oneill2011}
G.~C. {O'Neil}, P.~J. {Lowell}, J.~M. {Underwood}, and J.~N. {Ullom}.
\newblock Observations and modeling of large area
  normal-metal/insulator/superconductor refrigerator cooling from 300 mk to
  below 100 mk.
\newblock {\em ArXiv e-prints}, 1109.1273, Sep 2011.

\bibitem{Pekola2000}
J.~P. Pekola, D.~V. Anghel, T.~I. Suppula, J.~K. Suoknuuti, A.~J. Manninen, and
  M.~Manninen.
\newblock Trapping of quasiparticles of a nonequilibrium superconductor.
\newblock {\em Applied Physics Letters}, 76(19):2782--2784, 2000.

\bibitem{Court2008}
N.~A. Court, A.~J. Ferguson, Roman Lutchyn, and R.~G. Clark.
\newblock Quantitative study of quasiparticle traps using the
  single-cooper-pair transistor.
\newblock {\em Phys. Rev. B}, 77:100501, Mar 2008.

\bibitem{Rajauria2011}
S.~{Rajauria}, L.~M.~A. {Pascal}, P.~{Gandit}, F.~W.~J. {Hekking},
  B.~{Pannetier}, and H.~{Courtois}.
\newblock Efficient quasiparticle evacuation in superconducting devices.
\newblock {\em ArXiv e-prints}, 1106.4949, Jun 2011.

\bibitem{Bardas1995}
A.~Bardas and D.~Averin.
\newblock Peltier effect in normal-metal - superconductor microcontacts.
\newblock {\em Phys. Rev. B}, 52(17):12873--, Nov 1995.

\bibitem{Averin2008}
Dmitri~V. Averin and Jukka~P. Pekola.
\newblock Nonadiabatic charge pumping in a hybrid single-electron transistor.
\newblock {\em Phys. Rev. Lett.}, 101(6):066801, Aug 2008.

\bibitem{Hekking1993}
F.~W.~J. Hekking and Yu.~V. Nazarov.
\newblock Interference of two electrons entering a superconductor.
\newblock {\em Phys. Rev. Lett.}, 71(10):1625--1628, Sep 1993.

\bibitem{Hekking1994}
F.~W.~J. Hekking and Yu.~V. Nazarov.
\newblock Subgap conductivity of a superconductor--normal-metal tunnel
  interface.
\newblock {\em Phys. Rev. B}, 49(10):6847--6852, Mar 1994.

\bibitem{Vasenko2010}
A.~S. Vasenko, E.~V. Bezuglyi, H.~Courtois, and F.~W.~J. Hekking.
\newblock Electron cooling by diffusive normal metal - superconductor tunnel
  junctions.
\newblock {\em Phys. Rev. B}, 81:094513, Mar 2010.

\bibitem{Rajauria2008}
Sukumar Rajauria, P.~Gandit, T.~Fournier, F.~W.~J. Hekking, B.~Pannetier, and
  H.~Courtois.
\newblock Andreev current-induced dissipation in a hybrid superconducting
  tunnel junction.
\newblock {\em Phys. Rev. Lett.}, 100(20):207002, May 2008.

\bibitem{Lowell2011}
P.~J. {Lowell}, G.~C. {O'Neil}, J.~M. {Underwood}, and J.~N. {Ullom}.
\newblock Andreev reflections in micrometer-scale
  normal-insulator-superconductor tunnel junctions.
\newblock {\em ArXiv e-prints}, 1110.4839, Oct 2011.

\bibitem{SukumarThesis}
Sukumar Rajauria.
\newblock {\em Electronic refrigeration using superconducting tunnel
  junctions}.
\newblock PhD thesis, Universite Joseph Fourier, 2008.

\bibitem{Saira2007}
Olli-Pentti Saira, Matthias Meschke, Francesco Giazotto, Alexander~M. Savin,
  Mikko M\"ott\"onen, and Jukka~P. Pekola.
\newblock Heat transistor: Demonstration of gate-controlled electronic
  refrigeration.
\newblock {\em Phys. Rev. Lett.}, 99(2):027203, Jul 2007.

\bibitem{Pekola2008}
Jukka~P. Pekola, Juha~J. Vartiainen, Mikko M\"ott\"onen, Olli-Pentti Saira,
  Matthias Meschke, and Dmitri~V. Averin.
\newblock Hybrid single-electron transistor as a source of quantized electric
  current.
\newblock {\em Nat Phys}, 4(2):120--124, Feb 2008.

\bibitem{Pekola2007}
Jukka~P. Pekola, Francesco Giazotto, and Olli-Pentti Saira.
\newblock Radio-frequency single-electron refrigerator.
\newblock {\em Phys. Rev. Lett.}, 98(3):037201, Jan 2007.

\bibitem{Kafanov2009}
S.~Kafanov, A.~Kemppinen, Yu.~A. Pashkin, M.~Meschke, J.~S. Tsai, and J.~P.
  Pekola.
\newblock Single-electronic radio-frequency refrigerator.
\newblock {\em Phys. Rev. Lett.}, 103(12):120801, Sep 2009.

\bibitem{Pekola2007a}
J.~P. Pekola and F.~W.~J. Hekking.
\newblock Normal-metal-superconductor tunnel junction as a brownian
  refrigerator.
\newblock {\em Phys. Rev. Lett.}, 98(21):210604, May 2007.

\bibitem{VandenBroeck2006}
C.~Van~den Broeck and R.~Kawai.
\newblock Brownian refrigerator.
\newblock {\em Phys. Rev. Lett.}, 96:210601, Jun 2006.

\bibitem{Peltonen2011}
J.~T. Peltonen, M.~Helle, A.~V. Timofeev, P.~Solinas, F.~W.~J. Hekking, and
  J.~P. Pekola.
\newblock Brownian refrigeration by hybrid tunnel junctions.
\newblock {\em Phys. Rev. B}, 84:144505, Oct 2011.

\bibitem{Ingold1992}
Hermann Grabert and Michel~H. Devoret, editors.
\newblock {\em Single Charge Tunneling - Coulomb Blockade Phenomena in
  Nanostructures}, volume 294 of {\em NATO ASI Series B: Physics}.
\newblock Plenum Press, New York, 1992.

\bibitem{Stan2004}
Gheorghe Stan, Stuart~B. Field, and John~M. Martinis.
\newblock Critical field for complete vortex expulsion from narrow
  superconducting strips.
\newblock {\em Phys. Rev. Lett.}, 92:097003, Mar 2004.

\bibitem{Ullom1998a}
J.~N. Ullom, P.~A. Fisher, and M.~Nahum.
\newblock Magnetic field dependence of quasiparticle losses in a
  superconductor.
\newblock {\em Applied Physics Letters}, 73(17):2494--2496, 1998.

\bibitem{Peltonen2011a}
J.~T. {Peltonen}, J.~T. {Muhonen}, M.~{Meschke}, N.~B. {Kopnin}, and J.~P.
  {Pekola}.
\newblock Magnetic-field-induced stabilization of non-equilibrium
  superconductivity.
\newblock {\em ArXiv e-prints}, 1108.1544, Aug 2011.

\bibitem{Luukanen2000}
A.~Luukanen, M.~M. Leivo, J.~K. Suoknuuti, A.~J. Manninen, and J.~P. Pekola.
\newblock On-chip refrigeration by evaporation of hot electrons at sub-kelvin
  temperatures.
\newblock {\em Journal of Low Temperature Physics}, 120:281--290, 2000.

\bibitem{Clark2005}
A.~M. Clark, N.~A. Miller, A.~Williams, S.~T. Ruggiero, G.~C. Hilton, L.~R.
  Vale, J.~A. Beall, K.~D. Irwin, and J.~N. Ullom.
\newblock Cooling of bulk material by electron-tunneling refrigerators.
\newblock {\em Applied Physics Letters}, 86(17):173508, 2005.

\bibitem{Miller2008}
N.~A. Miller, G.~C. O'Neil, J.~A. Beall, G.~C. Hilton, K.~D. Irwin, D.~R.
  Schmidt, L.~R. Vale, and J.~N. Ullom.
\newblock High resolution x-ray transition-edge sensor cooled by tunnel
  junction refrigerators.
\newblock {\em Applied Physics Letters}, 92(16):163501, 2008.

\bibitem{OConnell2010}
A.~D. O'Connell, M.~Hofheinz, M.~Ansmann, Radoslaw~C. Bialczak, M.~Lenander,
  Erik Lucero, M.~Neeley, D.~Sank, H.~Wang, M.~Weides, J.~Wenner, John~M.
  Martinis, and A.~N. Cleland.
\newblock Quantum ground state and single-phonon control of a mechanical
  resonator.
\newblock {\em Nature}, 464(7289):697--703, Apr 2010.

\bibitem{Teufel2011}
J.~D. Teufel, T.~Donner, Dale Li, J.~W. Harlow, M.~S. Allman, K.~Cicak, A.~J.
  Sirois, J.~D. Whittaker, K.~W. Lehnert, and R.~W. Simmonds.
\newblock Sideband cooling of micromechanical motion to the quantum ground
  state.
\newblock {\em Nature}, 475(7356):359--363, July 2011.

\bibitem{Chan2011}
Jasper Chan, T.~P.~Mayer Alegre, Amir~H. Safavi-Naeini, Jeff~T. Hill, Alex
  Krause, Simon Groblacher, Markus Aspelmeyer, and Oskar Painter.
\newblock Laser cooling of a nanomechanical oscillator into its quantum ground
  state.
\newblock {\em Nature}, 478(7367):89--92, October 2011.

\bibitem{Hekking2008}
F.~W.~J. Hekking, A.~O. Niskanen, and J.~P. Pekola.
\newblock Electron-phonon coupling and longitudinal mechanical-mode cooling in
  a metallic nanowire.
\newblock {\em Phys. Rev. B}, 77(3):033401, Jan 2008.

\bibitem{Karvonen2007}
J.~T. Karvonen and I.~J. Maasilta.
\newblock Influence of phonon dimensionality on electron energy relaxation.
\newblock {\em Phys. Rev. Lett.}, 99(14):145503, Oct 2007.

\bibitem{Muhonen2009}
J.~T. Muhonen, A.~O. Niskanen, M.~Meschke, Yu.~A. Pashkin, J.~S. Tsai,
  L.~Sainiemi, S.~Franssila, and J.~P. Pekola.
\newblock Electronic cooling of a submicron-sized metallic beam.
\newblock {\em Applied Physics Letters}, 94(7):073101, 2009.

\bibitem{Sonne2010}
Gustav Sonne, Milton~E. Pe\~na Aza, Leonid~Y. Gorelik, Robert~I. Shekhter, and
  Mats Jonson.
\newblock Cooling of a suspended nanowire by an ac josephson current flow.
\newblock {\em Phys. Rev. Lett.}, 104(22):226802, Jun 2010.

\bibitem{Sonne2011}
Gustav Sonne and Leonid~Y. Gorelik.
\newblock Ground-state cooling of a suspended nanowire through inelastic
  macroscopic quantum tunneling in a current-biased josephson junction.
\newblock {\em Phys. Rev. Lett.}, 106(16):167205, Apr 2011.

\bibitem{Santandrea2011}
F.~Santandrea, L.~Y. Gorelik, R.~I. Shekhter, and M.~Jonson.
\newblock Cooling of nanomechanical resonators by thermally activated
  single-electron transport.
\newblock {\em Phys. Rev. Lett.}, 106(18):186803, May 2011.

\bibitem{Chamon2011}
Claudio Chamon, Eduardo~R. Mucciolo, Liliana Arrachea, and Rodrigo~B. Capaz.
\newblock Heat pumping in nanomechanical systems.
\newblock {\em Phys. Rev. Lett.}, 106(13):135504, Apr 2011.

\bibitem{Koppinen2009}
P.~J. Koppinen and I.~J. Maasilta.
\newblock Phonon cooling of nanomechanical beams with tunnel junctions.
\newblock {\em Phys. Rev. Lett.}, 102(16):165502, Apr 2009.

\bibitem{Zijlstra2007}
T.~Zijlstra, C.~F.~J. Lodewijk, N.~Vercruyssen, F.~D. Tichelaar, D.~N. Loudkov,
  and T.~M. Klapwijk.
\newblock Epitaxial aluminum nitride tunnel barriers grown by nitridation with
  a plasma source.
\newblock {\em Applied Physics Letters}, 91(23):233102, 2007.

\bibitem{Clark2004}
A.~M. Clark, A.~Williams, S.~T. Ruggiero, M.~L. van~den Berg, and J.~N. Ullom.
\newblock Practical electron-tunneling refrigerator.
\newblock {\em Applied Physics Letters}, 84(4):625--627, 2004.

\bibitem{Frank1997}
B.~Frank and W.~Krech.
\newblock Electronic cooling in superconducting tunnel junctions.
\newblock {\em Physics Letters A}, 235(3):281 -- 284, 1997.

\bibitem{Manninen1999}
A.~J. Manninen, J.~K. Suoknuuti, M.~M. Leivo, and J.~P. Pekola.
\newblock Cooling of a superconductor by quasiparticle tunneling.
\newblock {\em Appl. Phys. Lett.}, 74(20):3020--3022, May 1999.

\bibitem{Tirelli2008}
S.~Tirelli, A.~M. Savin, C.~Pascual Garcia, J.~P. Pekola, F.~Beltram, and
  F.~Giazotto.
\newblock Manipulation and generation of supercurrent in out-of-equilibrium
  josephson tunnel nanojunctions.
\newblock {\em Phys. Rev. Lett.}, 101(7):077004, Aug 2008.

\bibitem{Ferguson2008}
A.~J. Ferguson.
\newblock Quasiparticle cooling of a single cooper pair transistor.
\newblock {\em Applied Physics Letters}, 93(5):052501, 2008.

\bibitem{Tuominen1992}
M.~T. Tuominen, J.~M. Hergenrother, T.~S. Tighe, and M.~Tinkham.
\newblock Experimental evidence for parity-based 2\textit{e} periodicity in a
  superconducting single-electron tunneling transistor.
\newblock {\em Phys. Rev. Lett.}, 69:1997--2000, Sep 1992.

\bibitem{Quaranta2011}
O.~Quaranta, P.~Spathis, F.~Beltram, and F.~Giazotto.
\newblock Cooling electrons from 1 to 0.4 k with v-based nanorefrigerators.
\newblock {\em Applied Physics Letters}, 98(3):032501, 2011.

\bibitem{Savin2001}
A.~M. Savin, M.~Prunnila, P.~P. Kivinen, J.~P. Pekola, J.~Ahopelto, and A.~J.
  Manninen.
\newblock Efficient electronic cooling in heavily doped silicon by
  quasiparticle tunneling.
\newblock {\em Applied Physics Letters}, 79(10):1471--1473, 2001.

\bibitem{Savin2003}
Alexander Savin, Mika Prunnila, Jouni Ahopelto, Pasi Kivinen, P\"aivi
  T\"orm\"a, and Jukka Pekola.
\newblock Application of superconductor-semiconductor schottky barrier for
  electron cooling.
\newblock {\em Physica B: Condensed Matter}, 329-333(Part 2):1481 -- 1484,
  2003.
\newblock Proceedings of the 23rd International Conference on Low Temperature
  Physics.

\bibitem{Savin2004}
A~Savin, J~Pekola, M~Prunnila, J~Ahopelto, and P~Kivinen.
\newblock Electronic cooling and hot electron effects in heavily doped
  silicononinsulator film.
\newblock {\em Physica Scripta}, 2004(T114):57, 2004.

\bibitem{Buonomo2003}
B.~Buonomo, R.~Leoni, M.~G. Castellano, F.~Mattioli, G.~Torrioli, L.~Di.
  Gaspare, and F.~Evangelisti.
\newblock Electron thermometry and refrigeration with doped silicon and
  superconducting electrodes.
\newblock {\em Journal of Applied Physics}, 94:7784, 2003.

\bibitem{Kivinen2003}
P.~Kivinen, A.~Savin, M.~Zgirski, P.~T\"orm\"a, J.~Pekola, M.~Prunnila, and
  J.~Ahopelto.
\newblock Electron phonon heat transport and electronic thermal conductivity in
  heavily doped silicon-on-insulator film.
\newblock {\em Journal of Applied Physics}, 94(5):3201--3205, 2003.

\bibitem{Sergeev2005}
A.~Sergeev, M.~Yu. Reizer, and V.~Mitin.
\newblock Deformation electron-phonon coupling in disordered semiconductors and
  nanostructures.
\newblock {\em Phys. Rev. Lett.}, 94(13):136602, Apr 2005.

\bibitem{Prunnila2005}
M.~Prunnila, P.~Kivinen, A.~Savin, P.~T\"orm\"a, and J.~Ahopelto.
\newblock Intervalley-scattering-induced electron-phonon energy relaxation in
  many-valley semiconductors at low temperatures.
\newblock {\em Phys. Rev. Lett.}, 95(20):206602, Nov 2005.

\bibitem{Prunnila2007}
M.~Prunnila.
\newblock Electron--acoustic-phonon energy-loss rate in multicomponent electron
  systems with symmetric and asymmetric coupling constants.
\newblock {\em Phys. Rev. B}, 75(16):165322, Apr 2007.

\bibitem{Muhonen2011}
J.~T. Muhonen, M.~J. Prest, M.~Prunnila, D.~Gunnarsson, V.~A. Shah, A.~Dobbie,
  M.~Myronov, R.~J.~H. Morris, T.~E. Whall, E.~H.~C. Parker, and D.~R. Leadley.
\newblock Strain dependence of electron-phonon energy loss rate in many-valley
  semiconductors.
\newblock {\em Applied Physics Letters}, 98(18):182103, 2011.

\bibitem{Prest2011}
M.~J. {Prest}, J.~T. {Muhonen}, M.~{Prunnila}, D.~{Gunnarsson}, V.~A. {Shah},
  J.~S. {Richardson-Bulloc}, A.~{Dobbie}, M.~{Myronov}, R.~J.~H. {Morris},
  T.~E. {Whall}, E.~H.~C. {Parker}, and D.~R. {Leadley}.
\newblock Strain enhanced electron cooling in a degenerately doped
  semiconductor.
\newblock {\em ArXiv e-prints}, 1111.0465, Nov 2011.

\bibitem{Edwards1993}
H.~L. Edwards, Q.~Niu, and A.~L. de~Lozanne.
\newblock A quantum dot refrigerator.
\newblock {\em Applied Physics Letters}, 63(13):1815--1817, 1993.

\bibitem{Edwards1995}
H.~L. Edwards, Q.~Niu, G.~A. Georgakis, and A.~L. de~Lozanne.
\newblock Cryogenic cooling using tunneling structures with sharp energy
  features.
\newblock {\em Phys. Rev. B}, 52(8):5714--5736, Aug 1995.

\bibitem{Gasparinetti2011}
S.~Gasparinetti, F.~Deon, G.~Biasiol, L.~Sorba, F.~Beltram, and F.~Giazotto.
\newblock Probing the local temperature of a two-dimensional electron gas
  microdomain with a quantum dot: Measurement of electron-phonon interaction.
\newblock {\em Phys. Rev. B}, 83:201306, May 2011.

\bibitem{Giazotto2007}
Francesco Giazotto, Fabio Taddei, Michele Governale, Rosario Fazio, and Fabio
  Beltram.
\newblock Landau cooling in metal-semiconductor nanostructures.
\newblock {\em New Journal of Physics}, 9(12):439, 2007.

\bibitem{Pekola2000a}
J.~P. Pekola, A.~J. Manninen, M.~M. Leivo, K.~Arutyunov, J.~K. Suoknuuti, T.~I.
  Suppula, and B.~Collaudin.
\newblock Microrefrigeration by quasiparticle tunnelling in nis and sis
  junctions.
\newblock {\em Physica B: Condensed Matter}, 280(1-4):485 -- 490, 2000.

\bibitem{Nguyen2011}
H.~Q. {Nguyen}, L.~M.~A. {Pascal}, Z.~H. {Peng}, O.~{Buisson}, B.~{Gilles},
  C.~{Winkelmann}, and H.~{Courtois}.
\newblock {Etching suspended superconducting hybrid junctions from a
  multilayer}.
\newblock {\em ArXiv e-prints}, 1111.3541, Nov 2011.

\end{thebibliography}

\end{document}